\documentclass[aps,prd,twocolumn,amsmath,amssymb,amsfonts,nofootinbib,superscriptaddress,altaffilletter]{revtex4-1}

\usepackage{graphicx}
\usepackage{hyperref}
\hypersetup{
  bookmarksopen=true
}
\usepackage{amssymb}
\usepackage{amsmath}
\usepackage{color}
\usepackage{supertabular}
\usepackage{dcolumn}
\usepackage{multirow}

\def\beq{\begin{equation}}
\def\eeq{\end{equation}}
\def\bea{\begin{eqnarray}}
\def\eea{\end{eqnarray}}

\def\Tobs{T_{\textrm{\mbox{\tiny{obs}}}}}
\def\Tcoh{T_{\textrm{\mbox{\tiny{coh}}}}}

\def\ec{\textrm{~,}}

\def\fdot{\dot f}
\def\fddot{\ddot f}
\def\fdddot{\dddot f}
\def\ncur{n_{\rm c}}
\def\paramset{($f$,$\fdot$,$\fddot$)}
\def\Nfdot{N_{\dot f}}

\newif\ifshowfigs
\showfigstrue

\newcommand{\F}{2\mathcal{F}}
\newcommand{\FM}{2\mathcal{\bar F}}
\newcommand{\FMthresh}{2\mathcal{\bar {F}}_{\rm \!thresh}}
\newcommand{\FstatHz}{2\mathcal{\bar {F}}_0}
\newcommand{\Nseg}{N_{\rm seg}}
\newcommand{\mcoh}{m_{\rm coh}}
\newcommand{\msemicoh}{m_{\rm semi-coh}}

\newcommand{\fgwssb}{f}

\newcommand{\SNR}{\textrm{SNR}}

\newcommand{\Fstat}{$\mathcal{F}$-statistic}
\newcommand{\Fstatplusveto}{$\mathcal{F}^{+{\rm veto}}$-statistic}
\newcommand{\depth}{\mathcal{D}}
\newcommand{\depthc}{\mathcal{D}_{\rm Cas A}}
\newcommand{\depthv}{\mathcal{D}_{\rm Vela Jr}}
\newcommand{\psdwt}{{\bar S_h}}

\newcommand{\asdwtoff}{\sqrt{{\bar S_h(f)}}}

\newcommand{\casa}{Cas A}
\newcommand{\vela}{Vela Jr.}
\newcommand{\rmode}{$r$-mode}
\newcommand{\rmodes}{\rmode s}
\newcommand\T{\rule{0pt}{2.6ex}}       
\newcommand\B{\rule[-1.2ex]{0pt}{0pt}} 

\def\weave{{\footnotesize\textsc{Weave}}}

\def\fmax{976}

\def\fgw{f}
\def\frot{f_{\rm rot}}
\def\Izz{I_{\rm zz}}
\def\hsens{h_{\rm sens}^{95\%}}

\def\sci#1#2{#1\times10^{#2}}

\def\etal{{\it et al.}}
\def\vs{{\it vs.}}

\def\lowesthCasA{{\color{black}{$6.3\times10^{-26}$}}}
\def\lowesthVelaJr{{\color{black}{$5.6\times10^{-26}$}}}
\def\lowesthfreq{166}
\def\comment#1{{\color{black}#1}}

\begin{document}

\title{
  Search of the Early O3 LIGO Data
  for Continuous Gravitational Waves from the 
  Cassiopeia A and Vela Jr. Supernova Remnants
}

\let\mymaketitle\maketitle
\let\myauthor\author
\let\myaffiliation\affiliation
\author{LIGO Scientific Collaboration}
\author{Virgo Collaboration}

\date[\relax]{compiled \today}

\noaffiliation

\begin{abstract}
  We present directed searches for continuous gravitational waves from
  the neutron stars in the Cassiopeia A (\casa) and \vela\ supernova remnants.
  We carry out the searches in the LIGO detector data from the
  first six months of the third Advanced LIGO and Virgo observing run
  using the \weave\ semi-coherent method, which sums matched-filter detection-statistic values over
  many time segments spanning the observation period.
  No gravitational wave signal is detected in the search band of 20--\fmax\ Hz
  for assumed source ages greater than 300 years for \casa\
  and greater than 700 years for \vela\
  Estimates from simulated continuous wave
  signals indicate we achieve the most sensitive results to date across the
  explored parameter space volume, probing to strain magnitudes as
  low as $\sim$\lowesthCasA\ for \casa\ and $\sim$\lowesthVelaJr\ for Vela Jr.
  at frequencies near \lowesthfreq\ Hz at 95\%\ efficiency.
  
\end{abstract}

\maketitle

\section{Introduction}
\label{sec:introduction}

We report the results of the deepest search to date for continuous gravitational waves from the neutron stars
at the centers of the Cassiopeia A (\casa, G111.7$-$2.1)~\cite{bib:RyleSmith}
and Vela Jr. (G266.2$-$1.2)~\cite{bib:IyudinEtal} supernova remnants. \casa\ is just over 300 years old~\cite{bib:Hughes,bib:FesenEtal}, and
\vela\ may be as young as 700 years old~\cite{bib:IyudinEtal}. These extremely young objects have been the target of multiple searches
for continuous gravitational waves since
2010~\cite{bib:S5search,bib:S6search,bib:O1search,bib:O1AEI1,bib:O1AEI2,bib:O2viterbi,bib:O3aSNR}
because they may retain high rotation frequencies
and may possess appreciable non-axisymmetries from their recent births~\cite{bib:Ruderman,bib;BaymEtal,bib:PandharipandeEtal,bib:Zimmermann,bib:Cutler,bib:HaskellEtal,bib:FattoyevEtal,bib:SinghEtal,bib:MingEtal}.
Continuous emission
due to unstable \rmodes\ is also possible in such young stars~\cite{bib:Andersson,bib:Bildsten,bib:FriedmanMorsink,bib:OwenEtal,bib:Kojima}.

In this search, we analyze the first six months of data from the third observing run (O3a period) of the
Advanced Laser Interferometer Gravitational wave Observatory (Advanced LIGO \cite{bib:LIGO,bib:DetectorPaper}).
We achieve significantly improved sensitivity for \vela\ with respect to a recent O3a search using
a different method~\cite{bib:O3aSNR} and dramatically improved sensitivity for \casa\ with respect to previous searches of
O1, O2 and O3a LIGO and Virgo data~\cite{bib:S5search,bib:S6search,bib:O1search,bib:O1AEI1,bib:O1AEI2,bib:O2viterbi,bib:O3aSNR}.
The improvement with respect to similar, previous analyses of O1 data~\cite{bib:O1AEI1,bib:O1AEI2} comes
largely from the improved detector noise due to a variety of instrument upgrades~\cite{bib:O3performance},
including a ($\sim$3 db) improvement achieved with quantum squeezing~\cite{bib:O3squeezing}.

Given the immense pressure on its nuclear matter, one expects a neutron star to assume a
highly spherical shape in the limit of no rotation and, with rotation, to form
an axisymmetric oblate spheroid. A number of physical processes can disrupt the symmetry,
however, to produce quadrupolar gravitational waves from the stellar rotation.
Those processes include crustal distortions from cooling or accretion,
buried magnetic field energy and excitation of r-modes. Comprehensive reviews of
continuous gravitational wave emission mechanisms from neutron stars can be found in \cite{bib:Lasky_review,bib:GandG_review}

Central compact objects (CCOs) at the cores of
supernova remnants present interesting potential sources, especially those in remnants inferred from their
sizes and expansion rates to be young. Both the \casa\ and \vela\ remnants contain such objects, thought
to be young neutron stars. One can derive an estimated age-based upper
limit\footnote{This strain estimate  gives a rough benchmark upper limit
on what is possible in an optimistic scenario; its assumption that current rotation frequency
is small relative to the star's birth frequency becomes less plausible for the highest frequencies
searched in this analysis.}
on a CCO's continuous-wave strain amplitude by assuming the star's current rotation frequency is much lower than its rotation
frequency at birth and that the star's spin-down since birth has been dominated by gravitational wave
energy loss (``gravitar'' emission)~\cite{bib:cwcasamethod}:
\begin{equation}
h_{\rm age} =  (\sci{2.3}{-24})\left({1\>{\rm kpc}\over r}\right)\sqrt{\left({1000\>{\rm yr}\over\tau}\right)
\left({\Izz\over I_0}\right)}\>,
\label{eqn:agebasedlimit}
\end{equation}

\noindent where $r$ is the distance to the source, $\tau$ is its age and $\Izz$ is the star's moment of inertia
about its spin axis, with a fiducial value of $I_0 = 10^{38}$ kg$\cdot$ m$^2$.

\casa\ is perhaps the most promising example of a potential gravitational wave CCO source in a supernova remnant.
Its birth aftermath may have been observed by Flamsteed~\cite{bib:Hughes} $\sim$340 years ago in
1680, and the expansion of the visible shell is consistent
with that date~\cite{bib:FesenEtal}. Hence Cas~A, which is visible in X-rays~\cite{bib:Tananbaum,bib:HoEtal}
but shows no pulsations~\cite{bib:HalpernGotthelf}, is almost certainly a very young 
neutron star at a distance of about 3.3 kpc~\cite{bib:ReedEtal,bib:AlarieEtal}. From equation~\ref{eqn:agebasedlimit},
one finds an age-based strain limit of $\sim$$\sci{1.2}{-24}$, which is readily accessible to
LIGO and Virgo detectors in their most sensitive band.
       
The Vela~Jr. CCO is observed in X-rays~\cite{bib:PavlovEtal} and is potentially
quite close ($\sim$0.2 kpc) and young (690 yr)~\cite{bib:IyudinEtal}, for which one finds
a quite high age-based strain limit of $\sim$$\sci{1.4}{-23}$. Some prior continuous gravitational wave
searches have also conservatively assumed a more pessimistic distance ($\sim$1 kpc) and age (5100 yr), based on other measurements~\cite{bib:AllenEtalVelaJr}, for which the
age-based strain limit is $\sim$$\sci{1.0}{-24}$, still comparable to that of \casa.
As in the case of \casa, no pulsations have been detected from \vela~\cite{bib:KargaltsevEtal,bib:BeckerEtal}.

The remainder of this article is organized as follows: Section~\ref{sec:data} describes the data set used.
Section~\ref{sec:method} briefly describes the \weave\ search program~\cite{bib:WetteEtal} which uses
semi-coherent summing of a matched-filter detection statistic known as the \Fstat~\cite{bib:JKS}.
Section~\ref{sec:results} presents the results of the search.
Section~\ref{sec:sensitivity} discusses the method used to determine 95\%\ sensitivity as
an approximation to rigorous upper limits for bands in which all initial search outliers
have been followed up with more sensitive but computationally costly methods and dismissed
as not credible signals.
Section~\ref{sec:conclusions} concludes with a discussion of the results and
prospects for future searches.

\section{Data sets used}
\label{sec:data}

Advanced LIGO consists of two detectors, one in Hanford, Washington (designated H1),
and the other in Livingston, Louisiana (designated L1), separated by a $\sim$3000-km baseline~\cite{bib:LIGO}.
Each site hosts one, 4-km-long interferometer inside a vacuum envelope with the primary interferometer optics suspended by a
cascaded, quadruple suspension system,  affixed beneath an in-series pair of suspended optical tables,
in order to isolate them from external disturbances.
The interferometer mirrors act as test masses, and the passage of a gravitational wave induces a differential-arm length change
which is proportional to the gravitational-wave strain amplitude.

The third Advanced LIGO and Virgo data run (O3) began April 1, 2019 and ended March 27, 2020.
The first six months (April 1, 2019 to October 1, 2019),
prior to a 1-month commissioning break, is designated as the O3a period.
The analysis presented here uses only the O3a data set from the LIGO interferometers.  
The Virgo data has not been used in this analysis because of an unfavorable
tradeoff in computational cost for sensitivity gain, given the interferometer's higher noise level during the O3 run.
The systematic error in the amplitude calibration is estimated to be lower than 7\%\
(68\%\ confidence interval) for both LIGO detectors over all frequencies throughout O3a~\cite{bib:caliberror}. 

Prior to searching the O3a data for continuous wave (CW) signals, the quality of the data was assessed
and steps taken to mitigate the effects of instrumental artifacts. As in previous Advanced LIGO observing
runs~\cite{bib:linespaper}, instrumental ``lines'' (sharp peaks in fine-resolution, run-averaged H1 and L1 spectra)
are marked, and where
possible, their instrumental or environmental sources identified~\cite{bib:O3aLinelist}. The resulting database of artifacts proved helpful in
eliminating spurious signal candidates emerging from the search; no bands were vetoed {\it a priori}, however. In general,
the number of H1 lines in the O3a data was
similar to that observed in the O2 run, while the number of lines for L1 O3a data was substantially reduced.

As discussed in \cite{bib:O3aAllSky}, another type of artifact observed in the O3a data for both H1 and L1 were
relatively frequent and loud ``glitches'' (short, high-amplitude instrumental transients) with most of their spectral power
lying below $\sim$500 Hz.
To mitigate the effects of these glitches on O3a CW searches for signals below 475 Hz, a simple glitch-gating
algorithm was applied ~\cite{bib:gatingdoc,bib:detcharpaper} to excise the transients from the data.

\section{Analysis Method}
\label{sec:method}

This search relies upon semi-coherent averaging of \Fstat~\cite{bib:JKS} values computed for many short (several-day)
segments spanning nearly all of the O3a run period (2019 April 1 15:00 UTC -- 2019 October 1 15:00 UTC).
Section~\ref{sec:model} describes the signal model used in the analysis.
Section~\ref{sec:fstatistic} describes the mean \Fstat\ detection statistic at the core of the analysis.
Section~\ref{sec:weave_infrastructure} describes the \weave\ infrastructure for summing individual \Fstat\
values over the observation period, including the configuration choices for the searches presented in this article.
Section~\ref{sec:outlier_followup} describes the procedure used to follow up on
outliers found in the first stage of the hierarchical search.

\subsection{Signal model and parameter space searched}
\label{sec:model}

The signal templates assume a classical model of a spinning neutron star with a time-varying quadrupole moment that produces
circularly polarized gravitational radiation along the rotation axis, linearly polarized radiation
in the directions perpendicular to the rotation axis and elliptical polarization for the general case.
The strain signal model $h(t)$ for the source, as seen by the detector, is assumed to be the following function of time $t$:
\begin{eqnarray}
  h(t) & = & h_0\bigl(F_+(t, \alpha_0, \delta_0, \psi)\frac{1+\cos^2(\iota)}{2}\cos(\Phi(t))\vphantom{\frac{1+\cos^2(\iota)}{2}}\nonumber\\
  & & \quad\>+F_\times(t, \alpha_0, \delta_0, \psi)\cos(\iota)\sin(\Phi(t))\bigr)\ec
  \label{eqn:signalmodel}
\end{eqnarray}
\noindent
In Eq.~\ref{eqn:signalmodel}, $h_0$ is the intrinsic strain amplitude, $\Phi(t)$ is the signal phase,
$F_+$ and $F_\times$ characterize the detector responses to signals with ``$+$'' and ``$\times$''
quadrupolar polarizations \cite{bib:S4allsky}, and the sky location is described by right ascension $\alpha_0$ and
declination $\delta_0$.
In this equation, the star's orientation, which determines the polarization, is parametrized by the inclination angle
$\iota$ of its spin axis relative to the detector line-of-sight and by the angle $\psi$ of the axis projection on the plane of the sky.
The linear polarization case ($\iota=\pi/2$) is the most unfavorable 
because the gravitational wave flux impinging on the detectors
is smallest for an intrinsic strain amplitude $h_0$,
possessing eight times less incident strain power than
for circularly polarized waves ($\iota = 0,\>\pi$).

In a rotating triaxial ellipsoid model for a 
star at distance $r$ spinning at frequency $\frot$ about its (approximate) symmetry axis ($z$),
the amplitude $h_0$ can be expressed as
\begin{eqnarray}
  \label{eqn:hexpected}
h_0 & = & {4\,\pi^2G\epsilon\Izz{\fgw}^2\over c^4r}  \\
    & = & [1.1\times10^{-24}]\Bigl[{\epsilon\over10^{-6}}\Bigr]\Bigl[{\Izz\over I_0}\Bigr]\Bigl[{\fgw\over1\>{\rm kHz}}\Bigr]^2
\Bigl[{1\>{\rm kpc}\over r}\Bigr],
\end{eqnarray}
for which the gravitational radiation is emitted at frequency $\fgw=2\,\frot$.
The equatorial ellipticity $\epsilon$ is a useful, dimensionless measure of stellar non-axisymmetry:
\begin{equation}
  \label{eqn:ellipticity}
\epsilon \quad \equiv \quad {|I_{xx}-I_{yy}|\over I_{zz}}.
\end{equation}

Unstable \rmode\ emission~\cite{bib:Andersson,bib:Bildsten,bib:FriedmanMorsink,bib:OwenEtal,bib:Kojima} at gravitational wave frequency $\fgw$ (which for this model is $\sim$(4/3$)\frot$)
can be parametrized by a dimensionless amplitude $\alpha$
governing the strain amplitude~\cite{bib:Owen2010}:
\begin{equation}
  \label{eqn:hrmodes}
  h_0  = [3.6\times10^{-23}]\Bigl[{\alpha\over0.001}\Bigr]\Bigl[{\fgw\over1\>{\rm kHz}}\Bigr]^3
\Bigl[{1\>{\rm kpc}\over r}\Bigr].
\end{equation}

The phase evolution of the signal is given in the reference frame of the Solar System barycenter (SSB)
by the third-order approximation:
\begin{eqnarray}
\label{eqn:phase_evolution}
\Phi(t) & = & 2\pi\bigl(\fgwssb\cdot (t-t_0)+{1\over2}\fdot\cdot (t-t_0)^2 \nonumber \\
        &   & +{1\over6}\fddot\cdot (t-t_0)^3\bigr))+\phi_0\ec
\end{eqnarray}
where $\fgwssb$ is the SSB source frequency, $\fdot$ is the first frequency derivative (which, when negative,
is termed the spin-down), $\fddot$ is the second frequency derivative,
$t$ is the SSB time, and the initial phase $\phi_0$ is
computed relative to reference time $t_0$ (taken here to be the approximate mid-point of the O3a period: 2019 June 30 15:07:45 UTC -- GPS 1245942483).
When expressed as a function of the local time of ground-based detectors,
Eq.~\ref{eqn:phase_evolution} acquires sky-position-dependent Doppler shift terms~\cite{bib:JKS}.

In this analysis, we search a band of gravitational wave signal $f$ from 20 to \fmax\ Hz and a frequency derivative $\dot{f}$ range
governed by assumed minimum ages $\tau$ of each source. Detector noise deteriorates badly below 20 Hz because of ground motion,
and in the band around 1000 Hz because of resonant mechanical disturbances. Similar previous searches~\cite{bib:S5search,bib:S6search,bib:O1search}
have assumed a power law spin-down: $\fdot \propto -f^n$ with braking index $n$, with $n$
taking on values of 3 for magnetic dipole emission, 5 for GW quadrupole emission (gravitar) and 7 for \rmode\
emission. For a source that begins at a high frequency and spins down to a much lower present-day frequency
with a constant braking index, one expects $\fdot\approx {1\over n-1}(f/\tau)$. Allowing for $n$ to range
between 2 and 7 because of multiple potential spin-down contributions leads to the search range:
\begin{equation}
  -{f\over\tau} \le \fdot \le -{1\over6}{f\over\tau},
\end{equation}
which has been assumed in several previous searches~\cite{bib:S5search,bib:S6search,bib:O1search}. Here we take
a slightly more conservative approach, allowing the upper limit on $\fdot$ to reach zero, at modest additional computational cost,
while allowing for some time-dependent braking indices and uncertainties in the source's effective age.
The range in second frequency derivative
$\fddot$ is determined for any frequency $f$ and first derivative $\fdot$ by the same relation used in previous searches
(governed by the braking index range considered):
\begin{equation}
  2{\fdot^2\over f} \le \fddot \le 7{\fdot^2\over f}.
\end{equation}

Table~\ref{tab:locations} lists the maximum absolute values of $\fdot$ and $\fddot$ at the lowest
and highest search frequencies, along with the right ascensions and declinations used in the \casa\ and \vela\ searches.

\begin{table}[htb]
\begin{center}
  \begin{tabular}{lcc}\hline
    \T\B Source                                  &  Cassiopeia A~\cite{bib:CasASimbad}  &   Vela Jr.~\cite{bib:VelaJrSimbad}     \\
    \hline\hline
    \T\B Right ascension                              &  23h 23m 27.85s                      &   8h 52m 1.4s \\
    \T\B Declination                                  &  +58$^\circ$ 48' 42.8''               &   --46$^\circ$ 17' 53''  \\
    \hline
    \T\B Max. $|\fdot|$ (Hz/s) @20 Hz         &  $\sci{2.1}{-9}$                     &   $\sci{9.1}{-10}$  \\
    \T\B Max. $|\fdot|$ (Hz/s) @976 Hz        &  $\sci{1.0}{-7}$                     &   $\sci{4.4}{-8}$   \\
    \hline
    \T\B Max. $\fddot$ (Hz/s$^2$) @20 Hz     &  $\sci{1.6}{-18}$                    &   $\sci{2.9}{-19}$  \\
    \T\B Max. $\fddot$ (Hz/s$^2$) @976 Hz    &  $\sci{7.6}{-17}$                    &   $\sci{1.4}{-17}$  \\
\hline 
\end{tabular}
  \caption{Sky locations and maximum $|\fdot|$, $\fddot$ values used in the \casa\ and \vela\ searches at the lowest and highest frequencies.}
\label{tab:locations}
\end{center}
\end{table}

In searching this parameter space, we do not enforce a relation among ($f$,$\fdot$,$\fddot$), which means
that for an arbitrary combination, the implied {\it current} braking index $\ncur$, defined by $\ncur \equiv f\fddot/(\fdot)^2$, may take on arbitrarily large (unphysical) values. For a true power-law behavior over the observation period,
the implied third frequency derivative can be written $\fdddot = \ncur(2\ncur-1)(\fdot)^2/f^3$. In the initial search and
first two stages of outlier follow-up, the third derivative is taken to be zero, which is a good approximation for
braking indices below 7 for both sources. 

\subsection{The mean \Fstat}
\label{sec:fstatistic}

This search is based on a semi-coherent average of \Fstat\ values over many individual intervals of the 6-month
observing period. Within each segment of coherence time duration $\Tcoh$, the \Fstat~\cite{bib:JKS} is computed
as in previous searches, as a detection statistic proportional to the signal amplitude $h_0^2$,
maximized over $h_0$, the unknown orientation angles $\iota$ and $\psi$, and the phase constant $\phi_0$.
In Gaussian noise with no signal present,
the value of $\F$ follows a $\chi^2$ distribution with four degrees of freedom and has
an expectation value of four. The presence of a signal leads to a non-central $\chi^2$ distribution with
a non-centrality parameter proportional to $h_0^2\cdot\Tcoh$ and inversely proportional to the average power spectral density
of the detector noise. The non-centrality parameter also depends on the source's orientation and sky location,
and on the orientations and locations of the LIGO interferometers~\cite{bib:JKS}.

We compute a semi-coherent mean \Fstat\ we call $\FM$ from the average value of $\F$ over the $\Nseg$ segments into which the observing
period is divided:
\begin{equation}
  \FM = {1\over\Nseg}\sum_{i=1}^{\Nseg}\, \F_i. 
\end{equation}
In the absence of signal, this detection statistic too has an expectation value of four, but has the underlying shape of
a $\chi^2$ distribution with $4\Nseg$ degrees of freedom with a (rescaled) standard deviation of $\sqrt{8/\Nseg}$.
The presence of a signal leads to an offset in the mean that is approximately the same as the
non-centrality parameter above, for a fixed $\Tcoh$.

\subsection{The \weave\ infrastructure}
\label{sec:weave_infrastructure}

The \weave\ software infrastructure provides a systematic approach to covering the parameter space volume in
a templated search to ensure acceptable loss of signal-to-noise ratio (\SNR) for true signals lying between
template points~\cite{bib:WetteEtal}.
The \weave\ program combines together recent developments 
in template placement to use an optimal parameter-space metric~\cite{bib:WettePrix,bib:WetteMetric} and
optimal template lattices~\cite{bib:WetteLattice}. The package is versatile enough to be used in all-sky searches for unknown
sources. Here we use a simpler configuration applicable to well localized sources, such as \casa\ and \vela

In brief, a template grid in the parameter space is created for each time segment,
a grid that is appropriate to computing the
\Fstat\footnote{To understand better the effects of instrumental line artifacts,
in this initial exploration of the O3 data with the \weave\ method,
a ``pure'' \Fstat\ was used rather than the Bayesian-motivated \Fstatplusveto~\cite{bib:Fstatplusveto1,bib:Fstatplusveto2},
in which the \Fstat\ value is suppressed by the presence of line artifacts in one detector, but not in the other.} for a
coherence time $\Tcoh$ equal to the total observation period $\Tobs$ divided by $\Nseg$. The spacing of the
grid points in ($f$, $\fdot$, $\fddot$) is set according to a metric~\cite{bib:WettePrix,bib:WetteMetric} that ensures a worst-case
maximum mismatch $\mcoh$ defined by the fractional loss in $\F$ value due to a true signal not coinciding with a
search template.

Separately, a much finer grid is defined for the full observation period with respect to the midpoint of the
observation period, one with its own mismatch parameter $\msemicoh$, analogous to $\mcoh$, but defined
to be the average of the coherent mismatch values over all segments~\cite{bib:WetteMetric}.
It choice is set empirically in a tradeoff between sensitivity and computational cost. The \weave\
package creates at initialization a mapping between each point in the semi-coherent template grid and a
nearest corresponding point in each of the separate, coarser segment grids, accounting for frequency evolution.
The semi-coherent detection statistic
$\FM$ is constructed for each semi-coherent template from this mapping~\cite{bib:WetteEtal}. 

For the \casa\ and \vela\ searches presented here, a simulation study was carried out to evaluate tradeoffs
in achievable sensitivity for a small but diverse set of segment length choices ($\Tcoh$) and mismatch parameters $\mcoh$ and
$\msemicoh$, with a goal of staying within a maximum computational cost of $3\times10^6$
CPU core hours for the two searches combined, including for outlier follow-up ($\sim$10\%). Searching over only $f$ and $\fdot$ was also explored, but yielded poorer sensitivity.
In the end, we chose the \weave\
configuration parameters shown in Table~\ref{tab:configparameters}.

\begin{table*}
\begin{center}
  \begin{tabular}{lcccc}\hline
     Parameter                                        &  \hskip0.2in~\hskip0.2in &\casa\      & \hskip0.2in~\hskip0.2in & \vela\  \\
    \hline\hline
    \T\B Coherent mismatch $\mcoh$                    & \hskip0.2in~\hskip0.2in &  0.1             &  \hskip0.2in~\hskip0.2in &  0.1     \\
    \T\B Semi-coherent mismatch $\msemicoh$           & \hskip0.2in~\hskip0.2in &  0.2             &  \hskip0.2in~\hskip0.2in &  0.2     \\
    \T\B Coherence time (number of segments) for initial search   & \hskip0.2in~\hskip0.2in &  5.0 days (36)   & \hskip0.2in~\hskip0.2in & 7.5 days (24) \\
    \T\B Coherence time (number of segments) for 1st follow-up    & \hskip0.2in~\hskip0.2in &  10.0 days (18)  & \hskip0.2in~\hskip0.2in & 15.0 days (12) \\
    \T\B Coherence time (number of segments) for 2nd follow-up    & \hskip0.2in~\hskip0.2in &  20.0 days (9)   & \hskip0.2in~\hskip0.2in & 30.0 days (6)  \\
    \T\B Coherence time (number of segments) for 3rd follow-up    & \hskip0.2in~\hskip0.2in &  45.0 days (4)   & \hskip0.2in~\hskip0.2in & 60.0 days (3)  \\
    \hline 
\end{tabular}
  \caption{\weave\ configuration parameters used for the \casa\ and \vela\ searches.}
  \label{tab:configparameters}
\end{center}
\end{table*}

Search jobs are carried out in 0.1-Hz bands of $f$, with further divisions in $\fdot$, as needed, to keep
each job's computational duration between approximately 6 and 12 hours, for practical reasons.
Tables~\ref{tab:fdotdivisionsCasA}-\ref{tab:fdotdivisionsVelaJr} show the number of $\fdot$ divisions \vs\ frequency band for the two searches.

\def\tabpad{\quad\quad}
 
\begin{table}[htb]
\begin{center}
  \begin{tabular}{rcr}\hline
     \multicolumn{1}{c}{Frequency}  & \hskip0.2in~\hskip0.2in                         & Number of $\fdot$  \\
     \multicolumn{1}{c}{band}       & \hskip0.2in~\hskip0.2in                         & sub-ranges  \\
    \hline\hline
    \T\B 20--151 Hz \strut      & \hskip0.2in~\hskip0.2in              &  1  \tabpad \\
    \T\B 151--251 Hz \strut     & \hskip0.2in~\hskip0.2in               &  5  \tabpad \\
    \T\B 251--301 Hz \strut     & \hskip0.2in~\hskip0.2in               & 10  \tabpad \\
    \T\B 301--401 Hz \strut     & \hskip0.2in~\hskip0.2in               & 20  \tabpad \\
    \T\B 401--501 Hz \strut     & \hskip0.2in~\hskip0.2in               & 30  \tabpad \\
    \T\B 501--555 Hz \strut     & \hskip0.2in~\hskip0.2in               & 35  \tabpad \\
    \T\B 551--651 Hz \strut     & \hskip0.2in~\hskip0.2in               & 45  \tabpad \\
    \T\B 651--701 Hz \strut     & \hskip0.2in~\hskip0.2in               & 55  \tabpad \\
    \T\B 701--801 Hz \strut     & \hskip0.2in~\hskip0.2in               & 85  \tabpad \\
    \T\B 801--926 Hz \strut     & \hskip0.2in~\hskip0.2in               & 105 \tabpad \\
    \T\B 926--976 Hz \strut     & \hskip0.2in~\hskip0.2in               & 130 \tabpad \\
    \hline 
\end{tabular}
  \caption{Numbers of $\fdot$ sub-ranges into which the initial \casa\ search jobs (0.1-Hz sub-bands) are divided for different frequency search bands,
    in order to maintain job durations between about 6 and 12 computational hours. Each sub-band is subject to a 1000-candidate top-list.}
  \label{tab:fdotdivisionsCasA}
\end{center}
\end{table}

\begin{table}[htb]
\begin{center}
  \begin{tabular}{rcr}\hline
     \multicolumn{1}{c}{Frequency}  & \hskip0.2in~\hskip0.2in                         & Number of $\fdot$  \\
     \multicolumn{1}{c}{band}       & \hskip0.2in~\hskip0.2in                         & sub-ranges  \\
    \hline\hline
    \T\B 20--201 Hz \strut   & \hskip0.2in~\hskip0.2in                  &  1 \tabpad \\
    \T\B 201--401 Hz \strut  & \hskip0.2in~\hskip0.2in                  &  5 \tabpad \\
    \T\B 401--501 Hz \strut  & \hskip0.2in~\hskip0.2in                  & 10 \tabpad \\
    \T\B 501--701 Hz \strut  & \hskip0.2in~\hskip0.2in                  & 20 \tabpad \\
    \T\B 701--901 Hz \strut  & \hskip0.2in~\hskip0.2in                  & 40 \tabpad \\
    \T\B 901--976 Hz \strut  & \hskip0.2in~\hskip0.2in                  & 60 \tabpad \\
    \hline 
\end{tabular}
  \caption{Numbers of $\fdot$ sub-ranges into which the initial \vela\ search jobs (0.1-Hz sub-bands) are divided for different frequency search bands,
    in order to maintain job durations between about 6 and 12 computational hours. Each sub-band is subject to a 1000-candidate top-list.}
  \label{tab:fdotdivisionsVelaJr}
\end{center}
\end{table}

\subsection{Outlier follow-up}
\label{sec:outlier_followup}

Each individual job returns the \paramset\ values of the 1000 templates (``top-list'') with the largest (``loudest'')
$\FM$ values. For 0.1-Hz bands with $\Nfdot$ divisions in the $\fdot$ range, there are $\Nfdot\times1000$
values returned. Outlier templates to be followed up are those in these top-lists exceeding a frequency-dependent
threshold $\FMthresh(f)$ which rises slowly with $f$ as the number of distinct templates searched grows, thereby increasing
the statistical trials factor. A nominal threshold is set based on the signal-free $\chi^2$ distribution
with four degrees of freedom per segment such that the expectation value of outliers is one per 1-Hz band
in Gaussian noise, given the empirically obtained trials factor. Using the template
counts from the \weave\ configuration yields an empirical fitted function $\FMthresh(f) = \FstatHz f^a$, where
the parameters $\FstatHz$ and $a$ are listed in Table~\ref{tab:threshparameters}.

\begin{table}[htb]
\begin{center}
  \begin{tabular}{lcccc}\hline
    \T\B Source                         &  $\FstatHz$  &   $a$   & $\FMthresh($20 Hz$)$ & $\FMthresh($976 Hz$)$   \\
    \hline\hline
    \T\B Cassiopeia A \strut            &  7.64        &  0.0227      &  8.18    &    8.93  \\
    \hline
    \T\B Vela Jr.     \strut            &  8.48        &  0.027      &  9.19    &   10.21  \\
\hline 
\end{tabular}
  \caption{Parameters defining the analytic threshold function $\FMthresh(f) = \FstatHz f^a$ applied to the $\FM$ detection
    statistic to define initial outliers for follow-up. Threshold values evaluated for $f$ = 20 and 976 Hz are also shown.}
\label{tab:threshparameters}
\end{center}
\end{table}

In practice, non-Gaussian artifacts lead to much higher outlier counts in particular bands contaminated by
instrumental line sources (Sect.~\ref{sec:data}). In some cases strong instrumental lines can lead to more than 1000
templates from a single job that exceed the threshold for a particular 0.1-Hz band and range of $\fdot$ searched.
We refer to those cases as ``saturated'' since potentially interesting templates may be suppressed by the top-list cap.
Each of those cases is examined manually to assess instrumental contamination. Where such contamination is confirmed,
those bands are marked and excluded from those in which we quote strain sensitivities.
The appendix lists these 0.1-Hz bands.

For non-saturated sub-ranges of individual 0.1-Hz bands,
outliers exceeding the threshold $\FMthresh(f)$ are followed up in a sequential procedure where at each step, the coherence
time $\Tcoh$ is doubled (and hence the number of segments $\Nseg$ is halved). Because the non-centrality parameter
for the mean $\FM$ detection statistic scales approximately linearly with $\Tcoh$, one expects a nominal
doubling of the {\it excess} mean $\FM$ defined by $\FM-4$.

To be conservative and guided by simulations, we
require outliers passing a follow-up stage to display an increase of 60--70\%\ in
excess mean $\FM$ with respect to the previous stage, depending on source and follow-up stage.
Table~\ref{tab:excessincreases} lists the required increaes, which are lower for \casa\ than for \vela\ in
the first follow-up stages because its younger age leads to higher possible 3rd frequency derivatives which
are not searched over in those stages. 
The simulated signals used to guide these choices are nominally detectable but not loud, having
strain amplitudes ranging from $\sim$1.1--1.5 times the estimated strain amplitude $\hsens$ for which the
$\FMthresh(f)$ threshold yields 95\%\ efficiency (see Sect.~\ref{sec:sensitivity}). The required increases in
$\FM$ leads to an losses in overall signal efficiency below $\sim$2\% for braking indices below 7.
For each follow-up stage, the search space around each outlier's values of $f$, $\fdot$ and $\fddot$ was chosen to
be three times (in all dimensions) the template step sizes used in the previous stage.
In the third stage, the range of $\fdddot$ searches is from zero to twice the implied value of the 2nd-round
survivor, assuming a power law spindown during the observation period.
All of these follow-up requirements and resulting efficiencies were evaluated by end-to-end software injections. 

\begin{table}[htb]
\begin{center}
  \begin{tabular}{lccc}\hline
                                        &  Round 1   &  Round 2   & Round 3 \\
     Source                         &  increase  &  increase  & increase \\
    \hline\hline
    \T\B Cassiopeia A \strut            &  65\%\       &  60\%\      &  70\%\ \\
    \hline
    \T\B Vela Jr.     \strut            &  70\%\       &  70\%\      &  70\%\ \\
\hline 
\end{tabular}
  \caption{Required increases in excess mean $\FM$ in each stage of outlier follow-up. The third frequency derivative
    $\fdddot$ is taken to be zero in the first two stages, but explicitly searched over in the third follow-up stage.}
\label{tab:excessincreases}
\end{center}
\end{table}

In the first stage of follow-up, all outliers above threshold are evaluated. In that initial stage, which more finely samples the
parameter space, multiple outliers may survive the next threshold requirement. In successive stages,
only the loudest survivor corresponding to the outlier being evaluated is passed to the next stage of follow-up.
Pursuing only the loudest survivor per initial outlier preserves high detection efficiency for a true signal while reducing computational
cost from following up multiple candidate templates contaminated by the same instrumental disturbance. 

\section{Search results}
\label{sec:results}

The search described above was carried out on the O3a data for the \casa\ and \vela\ sources.
For \casa\ (\vela), there were $\sim$$\sci{2}{5}$ ($\sim$$\sci{1}{5}$) outliers above threshold from the initial search in bands
that were not excluded from consideration by severe instrumental artifacts.
These outliers were all followed up individually with a narrowed search and a doubling of the
coherence time. An outlier was considered to survive follow-up if the loudest candidate template
from its follow-up displayed an increase of 70\%\ or more in excess $\FM$ with respect to
the original outlier's excess $\FM$. This criterion 
led to O($\sci{2}{4}$) survivors for each source.
That loudest surviving template then served as a seed for a second
round of follow-up using another doubling of coherence time. Once again, survivors of the
round were defined by another increase by at least 70\%\ in excess $\FM$ with respect to
the seed template, leading to $\sim$$\sci{5}{3}$ ($\sim$$\sci{1}{3}$) 2nd-round survivors for \casa\ (\vela).

Survivors of this second round of follow-up were all clustered and the loudest template visually inspected, to assess
instrumental line contamination. Clustering was carried out in frequency using simple grouping of any survivor template
within 0.01 Hz of another survivor template.
Tables~\ref{tab:round2survivors_CasA}--\ref{tab:round2survivors_VelaJr} list the parameters of
the single loudest outlier in each cluster.
In nearly every band a loud instrumental artifact was apparent.
To identify these contaminations, we construct so-called ``strain histograms'' in which the summed power over the
observation period from a simulation of the nominal
signal candidate is superposed on a background estimate of the noise estimated via interpolation between neighboring
frequency bands. For computational efficiency, the summed power is approximated via a histogram of rescaled integer counts from each
30-minute digital Fourier transform used in the search. 
Except for signal templates with high-magnitude spin-downs, the histograms typically display at least one ``horn''
(narrow peak) from
an interval during the 6-month O3a period when the orbitally modulated frequency is relatively stationary.

\begin{table*}[htb]
\begin{center}
  \def\mc#1{\multicolumn{1}{c}{#1}}
\begin{tabular}{rrrcrrrcrrrcrrr}\hline
\mc{$f$} & \mc{$\dot f$} & \mc{$\ddot f$} & &\mc{$f$} & \mc{$\dot f$} & \mc{$\ddot f$} & &\mc{$f$} & \mc{$\dot f$} & \mc{$\ddot f$} & &\mc{$f$} & \mc{$\dot f$} & \mc{$\ddot f$} \\ 
\mc{(Hz)} & \mc{(nHz/s)} & \mc{(aHz/s$^2$)} & &\mc{(Hz)} & \mc{(nHz/s)} & \mc{(aHz/s$^2$)} & &\mc{(Hz)} & \mc{(nHz/s)} & \mc{(aHz/s$^2$)} & &\mc{(Hz)} & \mc{(nHz/s)} & \mc{(aHz/s$^2$)} \\ 
\hline
\hline
$22.2597$ & $-0.18$ & $85.6$
 & \hskip0.10in~\hskip0.10in & $64.8765$ & $-0.54$ & $24.5$
 & \hskip0.10in~\hskip0.10in & $515.1569$ & $-1.86$ & $20.2$
 & \hskip0.10in~\hskip0.10in & $630.3393$ & $-65.77$ & $-9.6$
\\
$22.6684$ & $-0.41$ & $50.4$
 & \hskip0.10in~\hskip0.10in & $64.9955$ & $-0.71$ & $3.7$
 & \hskip0.10in~\hskip0.10in & $520.4822$ & $-3.13$ & $4.3$
 & \hskip0.10in~\hskip0.10in & $630.3606$ & $-66.06$ & $30.6$
\\
$24.9982$ & $-0.27$ & $1.5$
 & \hskip0.10in~\hskip0.10in & $67.9953$ & $-0.74$ & $4.0$
 & \hskip0.10in~\hskip0.10in & *$520.6814$ & $-2.86$ & $-28.5$
 & \hskip0.10in~\hskip0.10in & $630.3788$ & $-64.06$ & $-4.5$
\\
$28.0001$ & $-0.13$ & $1.7$
 & \hskip0.10in~\hskip0.10in & $73.9948$ & $-0.82$ & $2.4$
 & \hskip0.10in~\hskip0.10in & *$520.7275$ & $-10.13$ & $47.5$
 & \hskip0.10in~\hskip0.10in & $630.6003$ & $-33.38$ & $20.4$
\\
$29.7984$ & $-0.33$ & $0.9$
 & \hskip0.10in~\hskip0.10in & $74.1949$ & $-0.81$ & $3.2$
 & \hskip0.10in~\hskip0.10in & *$520.7408$ & $-12.02$ & $7.6$
 & \hskip0.10in~\hskip0.10in & $630.7146$ & $-56.47$ & $13.4$
\\
$30.1978$ & $-0.38$ & $-5.5$
 & \hskip0.10in~\hskip0.10in & $74.9948$ & $-0.81$ & $4.9$
 & \hskip0.10in~\hskip0.10in & *$521.4661$ & $-0.62$ & $-6.4$
 & \hskip0.10in~\hskip0.10in & $638.3140$ & $-4.82$ & $0.8$
\\
$30.8993$ & $-0.11$ & $65.0$
 & \hskip0.10in~\hskip0.10in & $79.9947$ & $-0.78$ & $19.6$
 & \hskip0.10in~\hskip0.10in & *$521.4895$ & $-4.25$ & $17.9$
 & \hskip0.10in~\hskip0.10in & $640.4433$ & $-7.18$ & $-63.6$
\\
$30.9979$ & $-0.32$ & $4.1$
 & \hskip0.10in~\hskip0.10in & $84.3942$ & $-0.91$ & $4.0$
 & \hskip0.10in~\hskip0.10in & $521.5134$ & $-7.75$ & $-27.4$
 & \hskip0.10in~\hskip0.10in & $799.9366$ & $-8.06$ & $51.6$
\\
*$31.3903$ & $-0.33$ & $2.4$
 & \hskip0.10in~\hskip0.10in & $90.8932$ & $-1.11$ & $-12.7$
 & \hskip0.10in~\hskip0.10in & $521.5494$ & $-5.77$ & $21.0$
 & \hskip0.10in~\hskip0.10in & $898.7328$ & $-18.60$ & $135.3$
\\
$32.4983$ & $-0.10$ & $57.7$
 & \hskip0.10in~\hskip0.10in & $92.4500$ & $-0.21$ & $50.1$
 & \hskip0.10in~\hskip0.10in & *$522.1882$ & $-4.01$ & $1.0$
 & \hskip0.10in~\hskip0.10in & $898.9294$ & $-46.32$ & $36.2$
\\
$33.8982$ & $-0.10$ & $64.0$
 & \hskip0.10in~\hskip0.10in & $107.2843$ & $-0.33$ & $79.8$
 & \hskip0.10in~\hskip0.10in & $527.3181$ & $-5.26$ & $-10.7$
 & \hskip0.10in~\hskip0.10in & $898.9575$ & $-41.81$ & $58.2$
\\
$35.8831$ & $-0.13$ & $72.6$
 & \hskip0.10in~\hskip0.10in & $107.4488$ & $-1.15$ & $2.3$
 & \hskip0.10in~\hskip0.10in & $528.5258$ & $-3.66$ & $-5.6$
 & \hskip0.10in~\hskip0.10in & $898.9840$ & $-34.74$ & $56.2$
\\
$37.9276$ & $-0.43$ & $-28.1$
 & \hskip0.10in~\hskip0.10in & $119.9910$ & $-1.30$ & $6.1$
 & \hskip0.10in~\hskip0.10in & *$533.4444$ & $-5.94$ & $-0.1$
 & \hskip0.10in~\hskip0.10in & $898.9993$ & $-32.69$ & $58.5$
\\
$37.9974$ & $-0.41$ & $2.4$
 & \hskip0.10in~\hskip0.10in & $301.9999$ & $-28.76$ & $113.5$
 & \hskip0.10in~\hskip0.10in & $599.6389$ & $-53.61$ & $32.4$
 & \hskip0.10in~\hskip0.10in & $899.1333$ & $-41.93$ & $110.5$
\\
$38.8732$ & $-0.09$ & $70.0$
 & \hskip0.10in~\hskip0.10in & $316.7777$ & $-3.49$ & $12.2$
 & \hskip0.10in~\hskip0.10in & $599.9509$ & $-6.48$ & $0.7$
 & \hskip0.10in~\hskip0.10in & $899.1675$ & $-30.36$ & $107.9$
\\
$40.7971$ & $-0.45$ & $1.5$
 & \hskip0.10in~\hskip0.10in & $333.3861$ & $-27.40$ & $0.2$
 & \hskip0.10in~\hskip0.10in & $604.7859$ & $-56.67$ & $11.9$
 & \hskip0.10in~\hskip0.10in & $899.1826$ & $-24.69$ & $-31.6$
\\
$42.3823$ & $-1.10$ & $73.6$
 & \hskip0.10in~\hskip0.10in & $348.5643$ & $-5.62$ & $-49.1$
 & \hskip0.10in~\hskip0.10in & $606.4225$ & $-62.43$ & $30.0$
 & \hskip0.10in~\hskip0.10in & $899.2032$ & $-12.52$ & $17.5$
\\
$43.3672$ & $-1.14$ & $-30.9$
 & \hskip0.10in~\hskip0.10in & $368.5754$ & $-3.85$ & $32.5$
 & \hskip0.10in~\hskip0.10in & $606.7767$ & $-49.63$ & $9.0$
 & \hskip0.10in~\hskip0.10in & $906.6986$ & $-17.85$ & $120.9$
\\
$44.3524$ & $-1.18$ & $1.8$
 & \hskip0.10in~\hskip0.10in & $485.2598$ & $-9.74$ & $13.7$
 & \hskip0.10in~\hskip0.10in & $606.8898$ & $-52.72$ & $33.0$
 & \hskip0.10in~\hskip0.10in & $909.7752$ & $-49.35$ & $67.2$
\\
$45.3311$ & $-0.17$ & $66.3$
 & \hskip0.10in~\hskip0.10in & $485.2772$ & $-4.83$ & $2.7$
 & \hskip0.10in~\hskip0.10in & $612.2997$ & $-50.45$ & $34.4$
 & \hskip0.10in~\hskip0.10in & $910.1130$ & $-6.43$ & $-10.9$
\\
$46.5430$ & $-0.06$ & $30.6$
 & \hskip0.10in~\hskip0.10in & *$487.9915$ & $-4.36$ & $11.2$
 & \hskip0.10in~\hskip0.10in & $612.4531$ & $-54.58$ & $38.2$
 & \hskip0.10in~\hskip0.10in & $918.6993$ & $-15.63$ & $7.8$
\\
$46.7866$ & $-4.35$ & $9.6$
 & \hskip0.10in~\hskip0.10in & *$489.5465$ & $-12.82$ & $33.2$
 & \hskip0.10in~\hskip0.10in & $615.0071$ & $-57.32$ & $42.1$
 & \hskip0.10in~\hskip0.10in & $918.7291$ & $-27.25$ & $15.9$
\\
$51.6816$ & $-0.15$ & $97.0$
 & \hskip0.10in~\hskip0.10in & $493.7602$ & $-0.14$ & $104.8$
 & \hskip0.10in~\hskip0.10in & $615.0244$ & $-59.39$ & $19.3$
 & \hskip0.10in~\hskip0.10in & $918.8026$ & $-1.51$ & $1.9$
\\
$52.8052$ & $-0.85$ & $9.3$
 & \hskip0.10in~\hskip0.10in & $494.7315$ & $-2.90$ & $5.3$
 & \hskip0.10in~\hskip0.10in & $615.0373$ & $-61.00$ & $79.0$
 & \hskip0.10in~\hskip0.10in & $918.8164$ & $-24.96$ & $18.9$
\\
$54.7779$ & $-0.28$ & $-74.9$
 & \hskip0.10in~\hskip0.10in & $495.0989$ & $-0.03$ & $-4.9$
 & \hskip0.10in~\hskip0.10in & $615.0512$ & $-60.78$ & $74.8$
 & \hskip0.10in~\hskip0.10in & $922.5653$ & $-18.33$ & $7.3$
\\
$54.8952$ & $-0.55$ & $11.6$
 & \hskip0.10in~\hskip0.10in & $501.1989$ & $-47.07$ & $31.7$
 & \hskip0.10in~\hskip0.10in & $629.8061$ & $-10.30$ & $-9.6$
 & \hskip0.10in~\hskip0.10in & $922.5866$ & $-13.49$ & $-34.6$
\\
$57.0006$ & $-3.78$ & $-20.8$
 & \hskip0.10in~\hskip0.10in & $504.7002$ & $-3.39$ & $-1.6$
 & \hskip0.10in~\hskip0.10in & $629.8195$ & $-9.25$ & $2.2$
 & \hskip0.10in~\hskip0.10in & *$945.1703$ & $-19.85$ & $20.8$
\\
$60.7084$ & $-0.09$ & $8.9$
 & \hskip0.10in~\hskip0.10in & $506.4994$ & $-28.63$ & $20.5$
 & \hskip0.10in~\hskip0.10in & $629.8375$ & $-15.47$ & $16.1$
 & \hskip0.10in~\hskip0.10in & $945.1870$ & $-14.94$ & $-0.3$
\\
$60.8843$ & $-0.36$ & $59.6$
 & \hskip0.10in~\hskip0.10in & $511.9005$ & $-35.97$ & $-92.9$
 & \hskip0.10in~\hskip0.10in & $630.3149$ & $-64.10$ & $-1.9$
 & \hskip0.10in~\hskip0.10in & $945.5873$ & $-13.65$ & $-76.8$
\\
\hline
\end{tabular}

  \caption{Frequency parameters for the loudest \casa\ outlier in each cluster that survived round 2 follow-up.
    Outliers marked with asterisks were followed up with a third round.}
  \label{tab:round2survivors_CasA}
\end{center}
\end{table*}

\begin{figure*}[htbp]
  \ifshowfigs
  \includegraphics[width=6in]{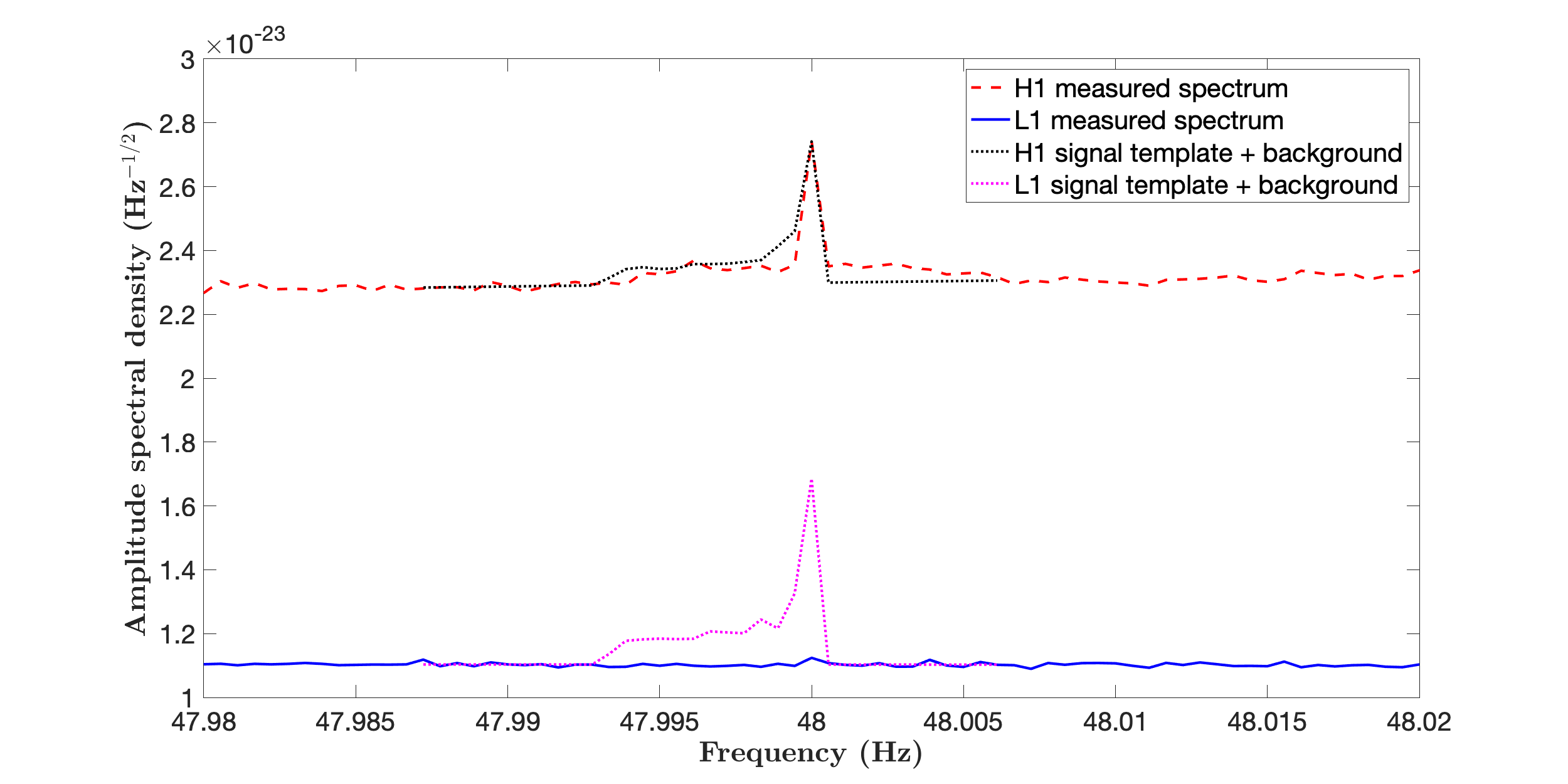}
  \includegraphics[width=6in]{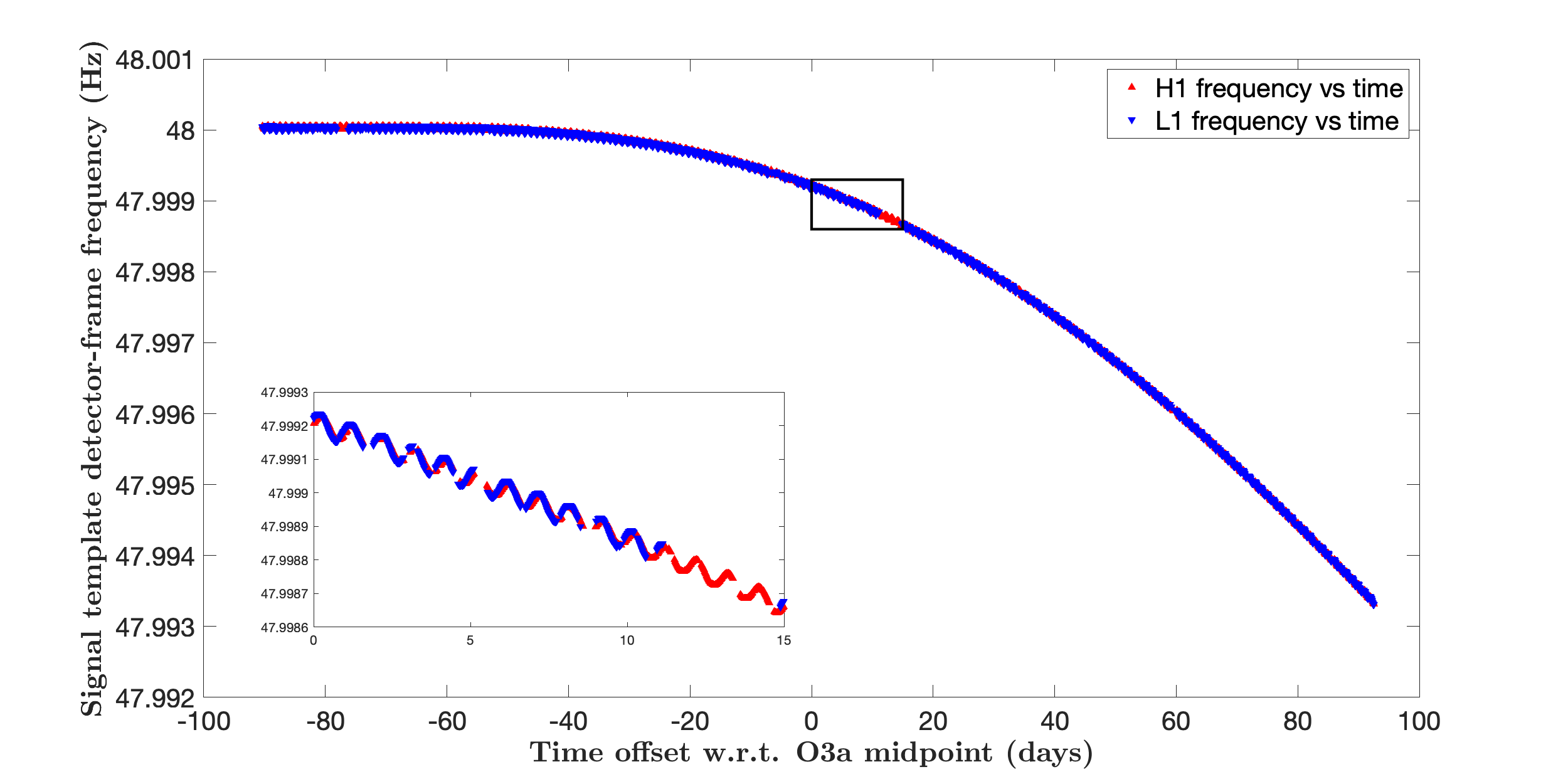}
  \fi
   \caption{{\it Upper panel:} Example of ``strain histogram'' graph for \casa\ used in vetoing outliers
    for which instrumental contamination is apparent. The  curves show the O3a-run-averaged
    H1 (red dashed) and L1 (blue solid) amplitude spectral densities in a narrow band containing an artifact at 48.000 Hz.
    The dotted curves show histograms of expected strain excess from H1 (black) and L1 (magenta)
    signal templates added to
    smooth backgrounds interpolated from neighboring frequency bands.
    In this depiction, the strain amplitude of the signal template has been
    magnified by an arbitrary factor large enough to make
    the signal's structure clear. The large excess power in the H1 data, not seen in the L1 data, despite
    comparable strain sensitivities and comparable sidereal-averaged antenna pattern sensitivities,
    excludes an astrophysical source for the H1 artifact. The fact that the artifact aligns in frequency
    with the putative signal's template peak in power confirms contamination of the outlier from an
    instrumental source. In addition, the line at precisely an integer frequency is part of a known instrumental spectral comb in
    the O3a H1 data. {\it Lower panel:} Graph of the corresponding template signal frequencies \vs\ time during O3a in the
    H1 and L1 interferometer reference frames, in which frequency points are plotted for only those 30-minute segments used in the analysis.
    One sees a relatively stationary period early in the run for \casa . The inset box shows a magnification of the frequency \vs\
    time graph for a 15-day period starting at the midpoint of the O3a interval, one that includes a multi-day period during which
    no data was collected from the L1 interferometer because Hurricane Barry disrupted observatory operations.
    The magnification makes more clear the
    diurnal modulation of the reference-frame frequency by the Earth's rotation about its axis, with slightly larger modulations
    seen for the lower-latitude L1 interferometer than for H1
    (color online).}
  \label{fig:strainhistogramCasA}
\end{figure*}

\begin{figure*}[htbp]
  \ifshowfigs
  \includegraphics[width=6in]{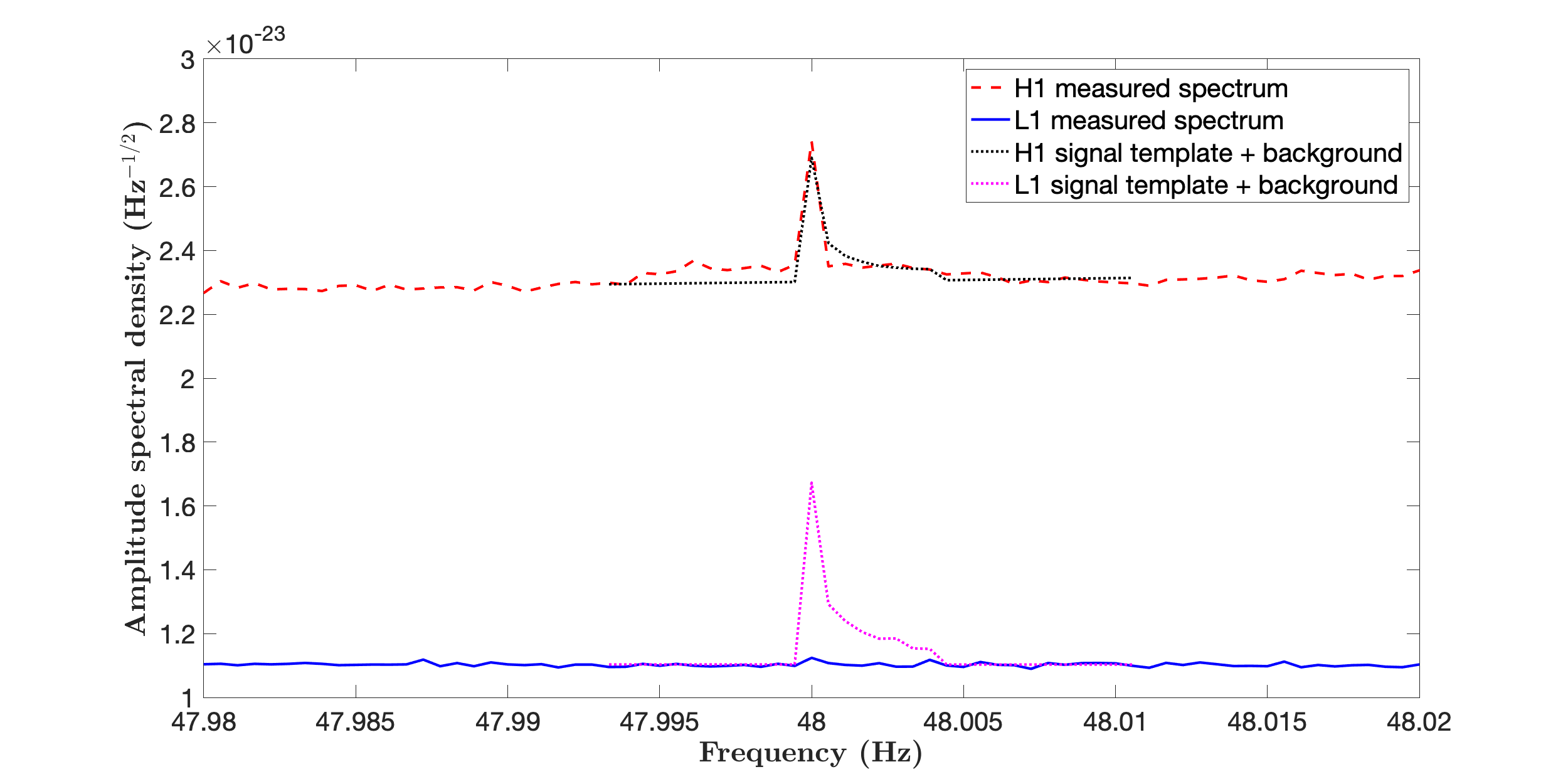}
  \includegraphics[width=6in]{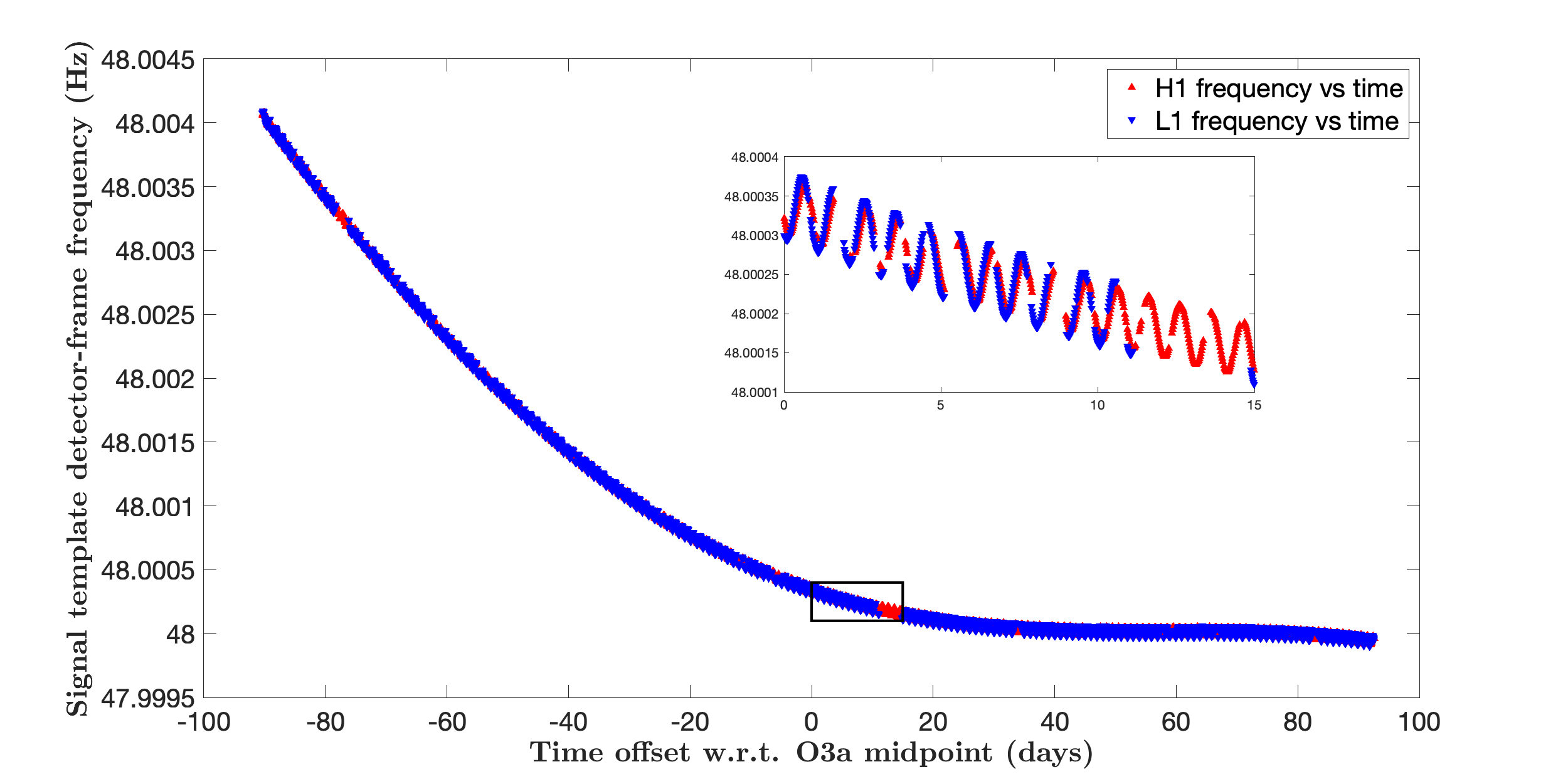}
  \fi
  \caption{Example of strain histogram and template frequency \vs\ time graphs for a \vela\ outlier with the same definitions (and colors) used for \casa\ in
    Fig.~\ref{fig:strainhistogramCasA}. One key difference with respect to \casa\ is that the interval of relatively stationary interferometer-frame
    frequency corresponding to the 48-Hz instrumental line occurs after the midpoint for \vela\ because of its
    different sky location from \casa\ (color online).}
  \label{fig:strainhistogramVelaJr}
\end{figure*}

\begin{table*}[htb]
\begin{center}
  \def\mc#1{\multicolumn{1}{c}{#1}}
\begin{tabular}{rrrcrrrcrrrcrrr}\hline
\mc{$f$} & \mc{$\dot f$} & \mc{$\ddot f$} & & \mc{$f$} & \mc{$\dot f$} & \mc{$\ddot f$} & & \mc{$f$} & \mc{$\dot f$} & \mc{$\ddot f$} & & \mc{$f$} & \mc{$\dot f$} & \mc{$\ddot f$} \\ 
\mc{(Hz)} & \mc{(nHz/s)} & \mc{(aHz/s$^2$)} & & \mc{(Hz)} & \mc{(nHz/s)} & \mc{(aHz/s$^2$)} & & \mc{(Hz)} & \mc{(nHz/s)} & \mc{(aHz/s$^2$)} & & \mc{(Hz)} & \mc{(nHz/s)} & \mc{(aHz/s$^2$)} \\ 
\hline
\hline
$20.0010$ & $-0.19$ & $-1.4$
 & \hskip0.10in~\hskip0.10in &$40.5022$ & $-0.40$ & $1.9$
 & \hskip0.10in~\hskip0.10in &$90.0002$ & $-0.35$ & $-3.0$
 & \hskip0.10in~\hskip0.10in &$612.1648$ & $-3.19$ & $2.9$
\\
$21.2899$ & $-0.10$ & $-26.8$
 & \hskip0.10in~\hskip0.10in &$40.7002$ & $-0.71$ & $-3.2$
 & \hskip0.10in~\hskip0.10in &$96.0010$ & $-1.12$ & $0.7$
 & \hskip0.10in~\hskip0.10in &$614.7653$ & $-3.77$ & $3.7$
\\
$22.2618$ & $-0.19$ & $-16.8$
 & \hskip0.10in~\hskip0.10in &$43.8618$ & $-0.43$ & $1.0$
 & \hskip0.10in~\hskip0.10in &$107.2955$ & $-1.01$ & $-8.8$
 & \hskip0.10in~\hskip0.10in &$629.8914$ & $-18.55$ & $6.2$
\\
$22.6708$ & $-0.53$ & $1.2$
 & \hskip0.10in~\hskip0.10in &$45.0022$ & $-0.35$ & $-16.4$
 & \hskip0.10in~\hskip0.10in &$130.9282$ & $-0.94$ & $-18.8$
 & \hskip0.10in~\hskip0.10in &$651.2010$ & $-7.28$ & $7.6$
\\
$23.6569$ & $-0.57$ & $-24.6$
 & \hskip0.10in~\hskip0.10in &$50.5949$ & $-0.44$ & $-5.9$
 & \hskip0.10in~\hskip0.10in &$299.3549$ & $-11.33$ & $3.0$
 & \hskip0.10in~\hskip0.10in &$652.8396$ & $-1.20$ & $-8.6$
\\
$24.6429$ & $-0.59$ & $0.3$
 & \hskip0.10in~\hskip0.10in &$51.1027$ & $-0.53$ & $6.9$
 & \hskip0.10in~\hskip0.10in &$487.2795$ & $-14.83$ & $26.2$
 & \hskip0.10in~\hskip0.10in &*$861.6319$ & $-9.66$ & $7.7$
\\
$25.9183$ & $-0.26$ & $2.4$
 & \hskip0.10in~\hskip0.10in &$52.8106$ & $-0.83$ & $-1.3$
 & \hskip0.10in~\hskip0.10in &*$488.2895$ & $-14.54$ & $7.3$
 & \hskip0.10in~\hskip0.10in &$898.8810$ & $-15.84$ & $0.4$
\\
$27.9117$ & $-0.25$ & $-3.4$
 & \hskip0.10in~\hskip0.10in &$53.8008$ & $-0.95$ & $-8.9$
 & \hskip0.10in~\hskip0.10in &$493.2190$ & $-7.06$ & $2.9$
 & \hskip0.10in~\hskip0.10in &$899.4151$ & $-27.55$ & $7.2$
\\
$28.9086$ & $-0.29$ & $1.9$
 & \hskip0.10in~\hskip0.10in &$54.7814$ & $-0.25$ & $0.8$
 & \hskip0.10in~\hskip0.10in &*$494.6662$ & $-22.17$ & $5.4$
 & \hskip0.10in~\hskip0.10in &$906.9044$ & $-14.44$ & $5.2$
\\
$29.8928$ & $-0.20$ & $3.3$
 & \hskip0.10in~\hskip0.10in &$57.0029$ & $-0.50$ & $-8.1$
 & \hskip0.10in~\hskip0.10in &$499.9158$ & $-4.49$ & $-21.5$
 & \hskip0.10in~\hskip0.10in &$910.0385$ & $-17.82$ & $12.3$
\\
$35.8866$ & $-0.24$ & $-21.4$
 & \hskip0.10in~\hskip0.10in &$60.7141$ & $-0.51$ & $-6.0$
 & \hskip0.10in~\hskip0.10in &$504.0999$ & $-12.64$ & $7.1$
 & \hskip0.10in~\hskip0.10in &$918.7510$ & $-16.75$ & $4.1$
\\
$37.5019$ & $-0.35$ & $-1.5$
 & \hskip0.10in~\hskip0.10in &$67.0034$ & $-0.63$ & $-0.9$
 & \hskip0.10in~\hskip0.10in &$510.9000$ & $-2.30$ & $13.1$
 & \hskip0.10in~\hskip0.10in &$918.8933$ & $-27.56$ & $3.5$
\\
$38.5020$ & $-0.37$ & $0.2$
 & \hskip0.10in~\hskip0.10in &$68.0035$ & $-0.66$ & $2.6$
 & \hskip0.10in~\hskip0.10in &$519.2834$ & $-14.17$ & $6.2$
 & \hskip0.10in~\hskip0.10in &$945.2994$ & $-7.15$ & $20.0$
\\
$38.8775$ & $-0.38$ & $0.9$
 & \hskip0.10in~\hskip0.10in &$73.4020$ & $-0.03$ & $-1.7$
 & \hskip0.10in~\hskip0.10in &$519.2962$ & $-10.62$ & $5.5$
 & \hskip0.10in~\hskip0.10in &$945.3565$ & $-17.64$ & $1.7$
\\
$38.9301$ & $-0.07$ & $-28.6$
 & \hskip0.10in~\hskip0.10in &$74.6306$ & $-0.70$ & $0.5$
 & \hskip0.10in~\hskip0.10in &$520.5149$ & $-5.31$ & $14.3$
 & \hskip0.10in~\hskip0.10in &\\
$39.8743$ & $-0.39$ & $1.5$
 & \hskip0.10in~\hskip0.10in &$85.6896$ & $-1.06$ & $-3.7$
 & \hskip0.10in~\hskip0.10in &$521.6642$ & $-18.22$ & $3.0$
 & \hskip0.10in~\hskip0.10in &\\
\hline
\end{tabular}

  \caption{Frequency parameters for the loudest \vela\ outlier in each cluster that survived round 2 follow-up.
    Outliers marked with asterisks were followed up with a third round.}
  \label{tab:round2survivors_VelaJr}
\end{center}
\end{table*}
We discard outliers for which the signal template's shape either aligns with a spectral artifact known to be instrumental, or else appears much louder in one detector than the other which is inconsistent with time-averaged antenna pattern sensitivities.
Figures~\ref{fig:strainhistogramCasA}-\ref{fig:strainhistogramVelaJr}
show example strain histograms for \casa\ and \vela\ outliers that are both heavily contaminated by an H1 spectral line at 48.000 Hz.
Figures~\ref{fig:strainhistogramCasA}-\ref{fig:strainhistogramVelaJr} also 
show graphs of the outlier templates' detector-frame frequencies \vs\ time during the O3a period, illustrating
periods of relatively stationary frequency. For these templates there is an approximate cancellation between
the source intrinsic spin-down and an apparent spin-up of frequency caused by the Earth's general acceleration
toward the direction of \casa\ early in O3a and toward the direction of \vela\ after
the midpoint of O3a. Since source frequencies are defined by the midpoint of
the O3a run, the \casa\ (\vela) template frequencies susceptible to this stationarity generally lie below (above) the detector-frame
frequency of the line artifact contaminating the template recovery.

A small number of outlier clusters for which a sharp line contamination is not the obvious cause were examined further.
The \casa\ outliers at 52.8052 Hz and 145.3899 Hz (in a saturated sub-band) are due to contamination from loud ``hardware injections.''
These injections are simulated signals imposed via modulated forces on interferometer mirrors during data taking. See \cite{bib:O3aAllSky,bib:O1hwinjections}
and \cite{bib:GWOSC} for more details on the hardware injections carried out during the O3a run.
For these outliers, the contaminations arise from injection ``Inj5''  and ``Inj6'' (see Table IV of \cite{bib:O3aAllSky}),
which both simulate CW sources near the sky location of \casa. The Inj5 injection is loud enough to show up as a \vela\ outlier too.


The 11 \casa\ outliers in Table~\ref{tab:round2survivors_CasA} and 3 \vela\ outliers in
Table~\ref{tab:round2survivors_VelaJr} that are marked with asterisks occur in spectral bands
in which instrumental lines are plentiful, but for which no clearcut artifact allows
immediate discarding of the outliers. For these outliers, a third round of follow-up was carried out,
with a third increase in coherence time:
from 20 to 45 days (from 9 to 4 segments) for \casa\ and from 30 to 60 days (from 6 to 3 segments) for \vela.
Because of the lengths of these coherently analyzed segments, these follow-ups included a search over the
third frequency derivative $\fdddot$. Simulations indicate that a 70\%\ increase in excess $\FM$ is
a conservative requirement, including for unphysically large braking indices, given that preceding follow-up stages
do not allow for a non-zero $\fdddot$. 
None of these outliers satisfies this 70\%\ requirement, and none has a braking index in the range 1--7.

\begin{figure*}[htbp]
  \ifshowfigs
  \includegraphics[width=7.in]{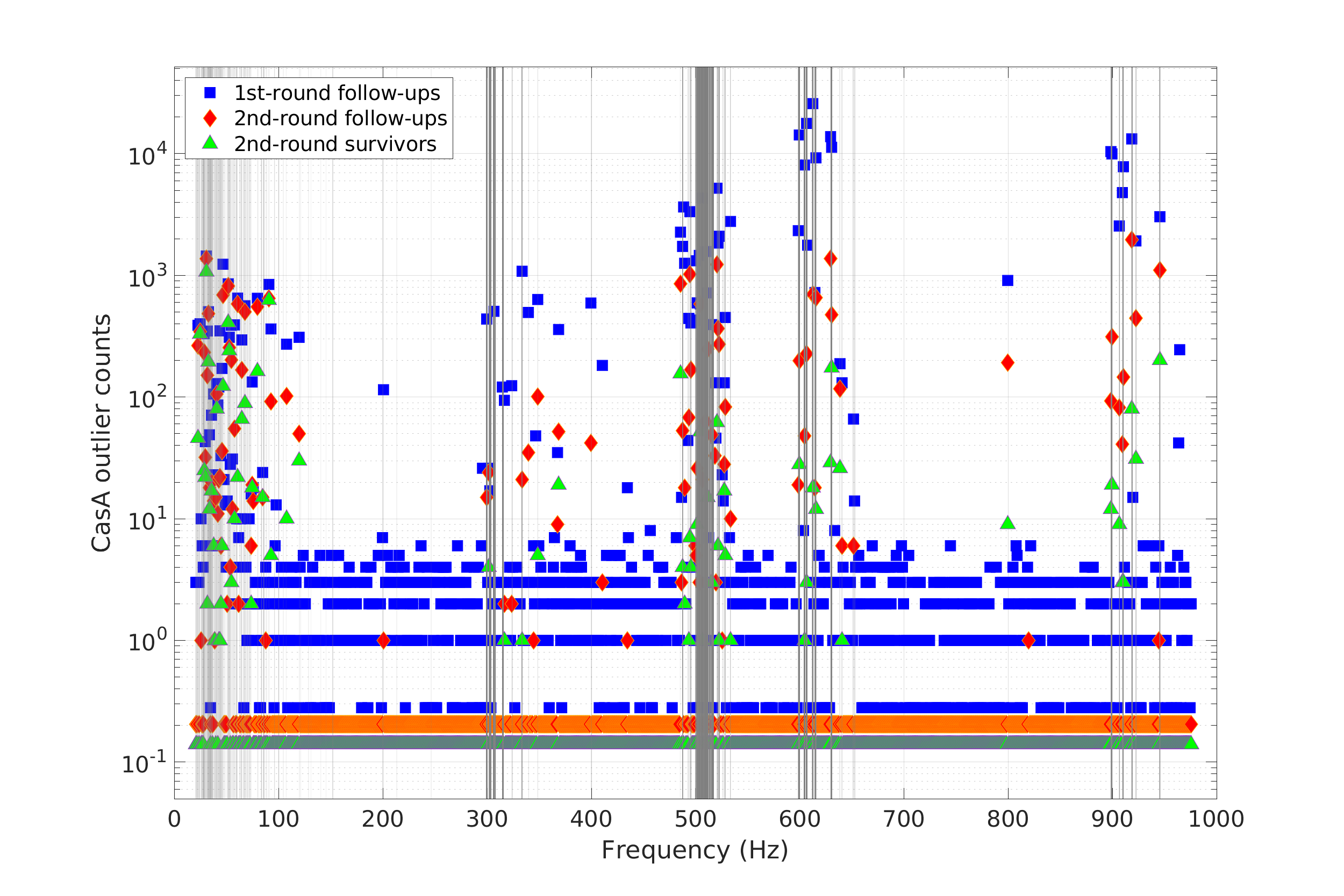}
  \fi
  \caption{Counts \vs\ frequency in 1-Hz bins for the initial \casa\ search outliers (blue squares), 1st-round follow-up survivors (red diamonds)
    and 2nd-round follow-up survivors (green triangles). The vertical gray bands denote consolidated 0.1-Hz sub-bands displaying saturation
    in the initial search. One sees high outlier counts and saturations primarily at low frequencies, near test-mass violin modes (resonant
    vibration modes of silica fibers around 500 Hz)
    and at harmonics of beam-splitter violin modes (above 300 Hz and near-integer multiples).
    Counts equal to zero for different stages are depicted on the
    vertical logarithmic scale by distinct fractions less than one
    (color online).}
  \label{fig:CasAoutliercounts}
\end{figure*}

\begin{figure*}[htbp]
  \ifshowfigs
  \includegraphics[width=7.in]{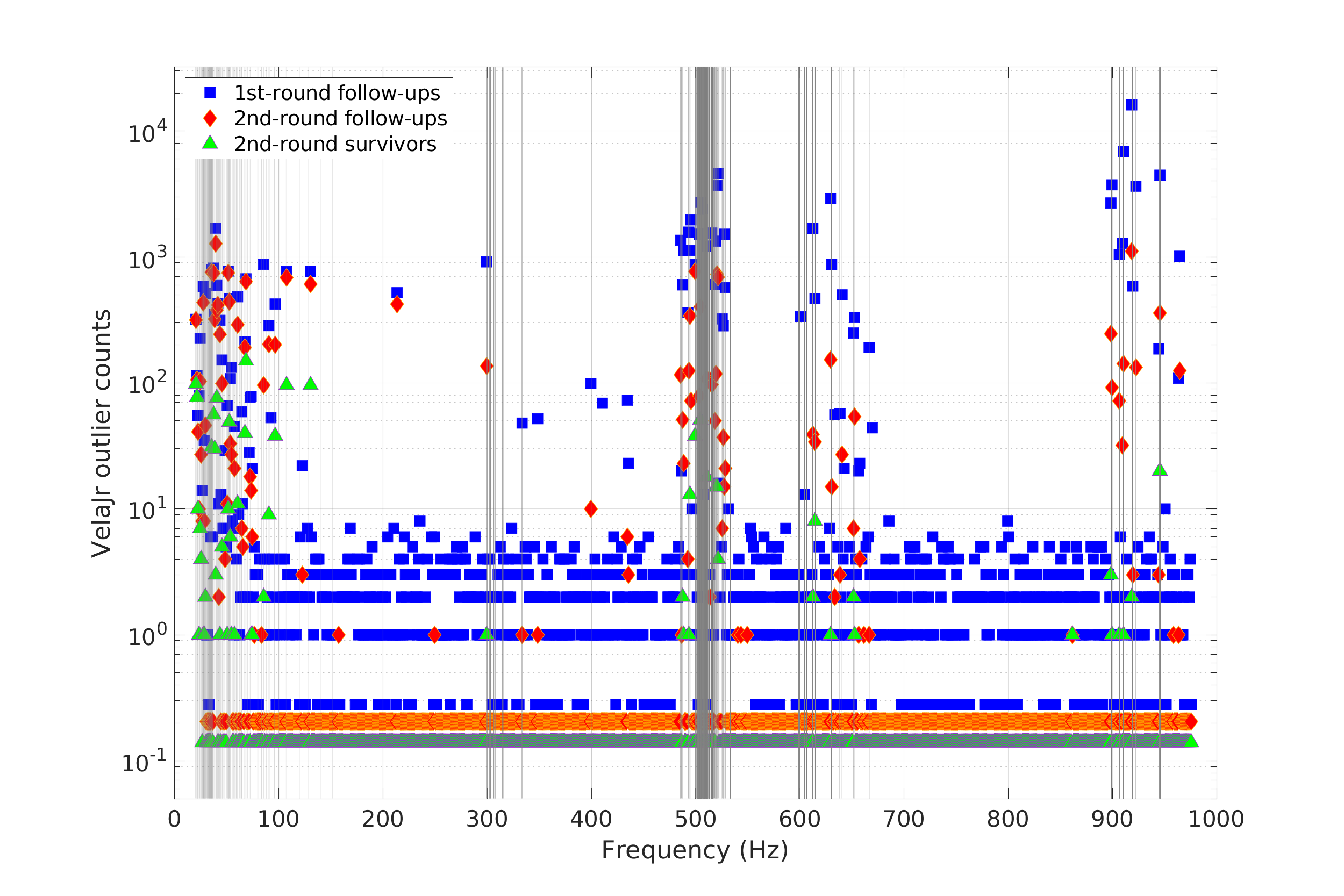}
  \fi 
  \caption{Counts \vs\ frequency in 1-Hz bins for the initial \vela\ search outliers (blue squares), 1st-round follow-up survivors (red diamonds)
    and 2nd-round follow-up survivors (green triangles). The vertical gray bands denote consolidated 0.1-Hz sub-bands,
    as in Fig.~\ref{fig:CasAoutliercounts}
    Counts equal to zero for different stages are depicted on the
    vertical logarithmic scale by distinct fractions less than one
    (color online).}
  \label{fig:VelaJroutliercounts}
\end{figure*}

Figures~\ref{fig:CasAoutliercounts}-\ref{fig:VelaJroutliercounts} show the \casa\ and \vela\  outlier and survivor counts
in 1-Hz bands for the multiple stages of analysis, starting with outliers exceeding the threshold $\FMthresh(f)$ and
proceeding to those surviving the successive requirements that the excess $\FM$ increase by 70\%\ each round of follow-up.
Saturated sub-bands listed in the appendix are shaded. 

We conclude that there is no significant evidence in this analysis for a continuous wave signal from
the compact objects at the centers of the \casa\ or \vela\ supernova remnants.

\section{Estimating search sensitivity}
\label{sec:sensitivity}

Given the absence of a detection, we quote 95\%-efficiency amplitude sensitivities $\hsens$
for every band in which there were no outliers above the initial $\FM$ threshold or for which every outlier was
followed up and found not to be a credible signal.  Those bands (listed in the appendix) with
at least one $\fdot$ interval exhibiting a saturated candidate top-list are excluded from the sensitivities
presented here.

\begin{figure}[htbp]
  \ifshowfigs
  \includegraphics[width=3.5in]{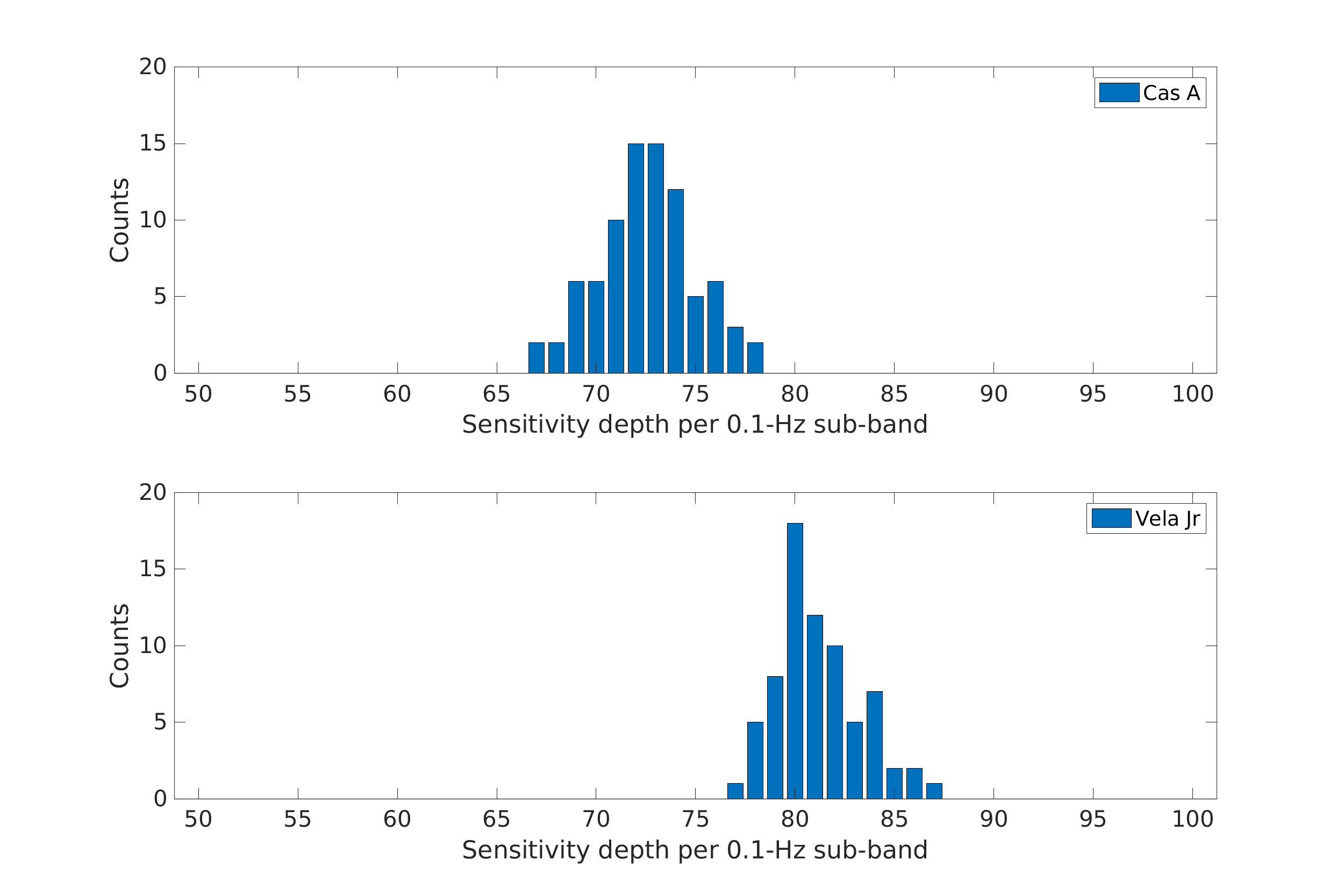}
  \fi
  \caption{Aggregated distributions of sensitivity depths (Eq.~\ref{eqn:sensitivitydepth}) for \casa\ (upper) and \vela\ (lower)
    based on 84 and 71 samples, respectively, of 0.1-Hz search sub-bands spanning the full 20--976 search band.
    The widths of the distributions are dominated by the depth variation with respect to frequency, which
    we empirically fit to a linear function of negative slope.}
  \label{fig:sensitivitydepth}
\end{figure}

We quote $\hsens$ values rather than
rigorous 95\% confidence level upper limits, in order to reduce computational cost.
To determine the sensitivity estimates, we use simulated signal injections to
perform rigorous upper limit determination for a sampling of 0.1-Hz frequency bands
(1000 injections per 0.1-Hz band) distributed over the search range; 84 bands were sampled for \casa, 71 for \vela\
Each upper limit is derived from a signal amplitude $\hsens$
that gives 95\%\ detection efficiency for a loudest $\FM$ value equal
to $\FMthresh(f)$ (given that all outliers above this threshold have been followed up and
eliminated). The sampled upper limits are used to determine an approximate scale factor between nominal detector sensitivity
and upper limit sensitivity for a given 0.1-Hz band, known as sensitivity depth $\depth$~\cite{bib:sensitivitydepth}:
\begin{equation}
  \depth(f) \equiv {\asdwtoff\over \hsens},
  \label{eqn:sensitivitydepth}
\end{equation}
\noindent where $\asdwtoff$ is an estimate of the effective strain amplitude spectral noise density. For non-stationary
detector noise, we use an inverse-noise weighted estimate for each frequency bin $j$ from the two interferometers:
\begin{equation}
  \psdwt(f_j) = {\sum_i w_{ij} S_h(f_j)\over\sum_i w_{ij}},
  \label{eqn:weightedasd}
\end{equation}
where $i$ ranges over Fourier transforms of 30-minute segments of the H1 and L1 data, and
$w_{ij}$ is a weight equal to the average inverse power spectral density for 50 neighboring frequency bins $j'\ne j$
in the same Fourier transform $i$:
\begin{equation}
  w_{ij} \equiv {1\over50} \sum_{j'}{1\over S_h(f_{j'})}
\end{equation}
\noindent for $|j'-j|\le25$ and $j'\ne j$. This weighting de-emphasizes noisy segments of data, similarly to the weighting
used to define the \Fstat. Figure~\ref{fig:sensitivitydepth} shows the full distributions in the resulting sensitivity
depths for \casa\ and \vela\ over the span of the search space, including significant spread from a slow decline
in depth with increasing frequency due to the higher threshold $\FMthresh(f)$.
From simple linear fits to depth \vs\ frequency, we determine
frequency-dependent scale factors which have values at 500 Hz of
$\depthc$ = \comment{72.4} Hz$^{-{1\over2}}$  and $\depthv$ = \comment{81.2} Hz$^{-{1\over2}}$ 
with slopes of $\sci{-4.8}{-3}$ Hz$^{-{3\over2}}$ and  $\sci{-5.6}{-3}$ Hz$^{-{3\over2}}$, respectively. 
The ratio of depths $\depthv$/$\depthc$ = \comment{1.12$\pm$0.01} at 500 Hz is consistent with the approximate expected
ratio of [(7.5 days)/(5.0 days)]$^{1/4}$ = 1.11 for these semi-coherent searches.

\section{Conclusions}
\label{sec:conclusions}

We have performed the deepest search to date for continuous gravitational waves from
compact stars in the centers of the Cassiopeia A and
Vela Jr. supernova remnants.
Our search yieded no detections.

\begin{figure*}[htbp]
  \ifshowfigs
  \includegraphics[width=7.0in]{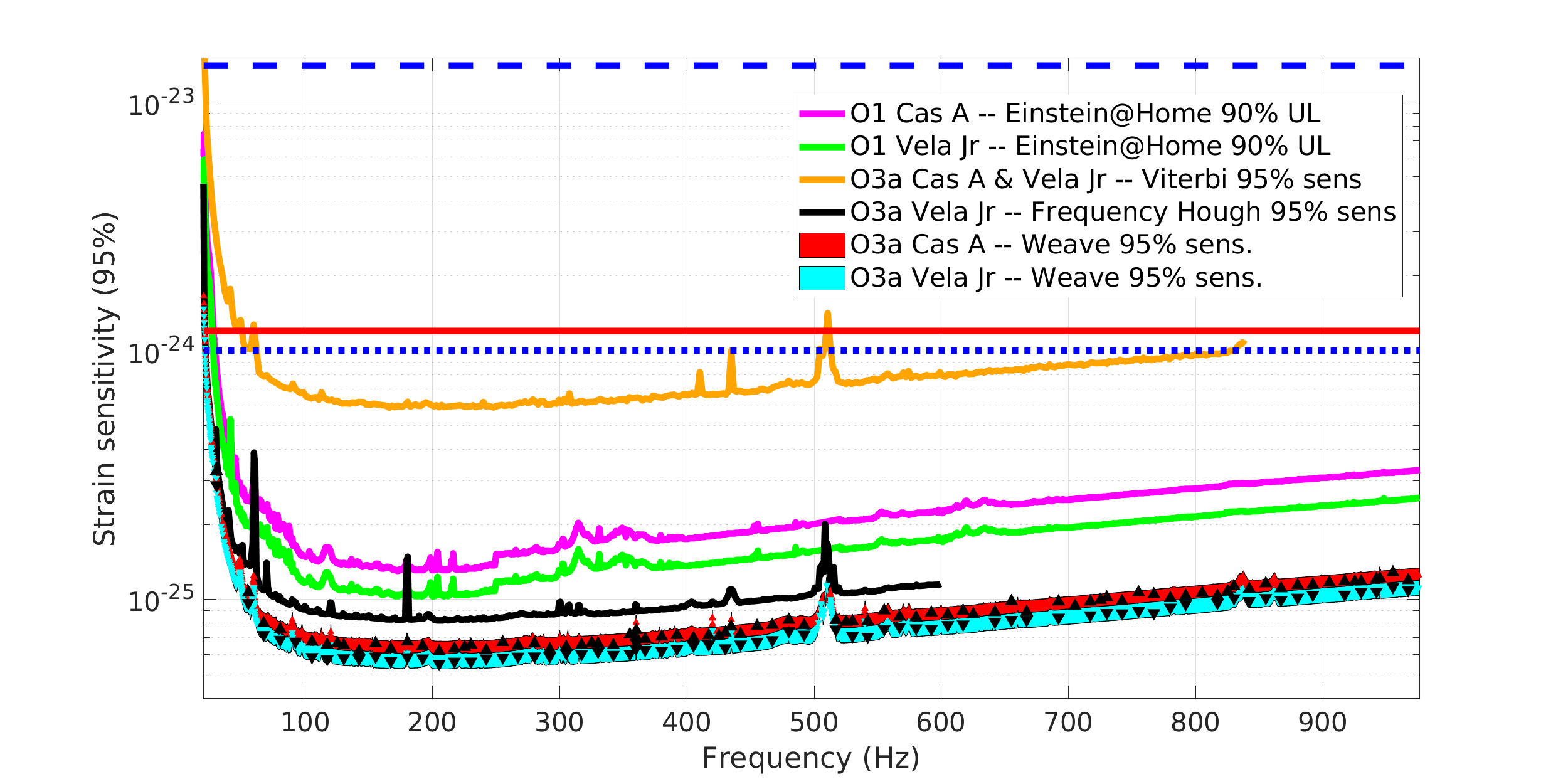}
  \includegraphics[width=7.0in]{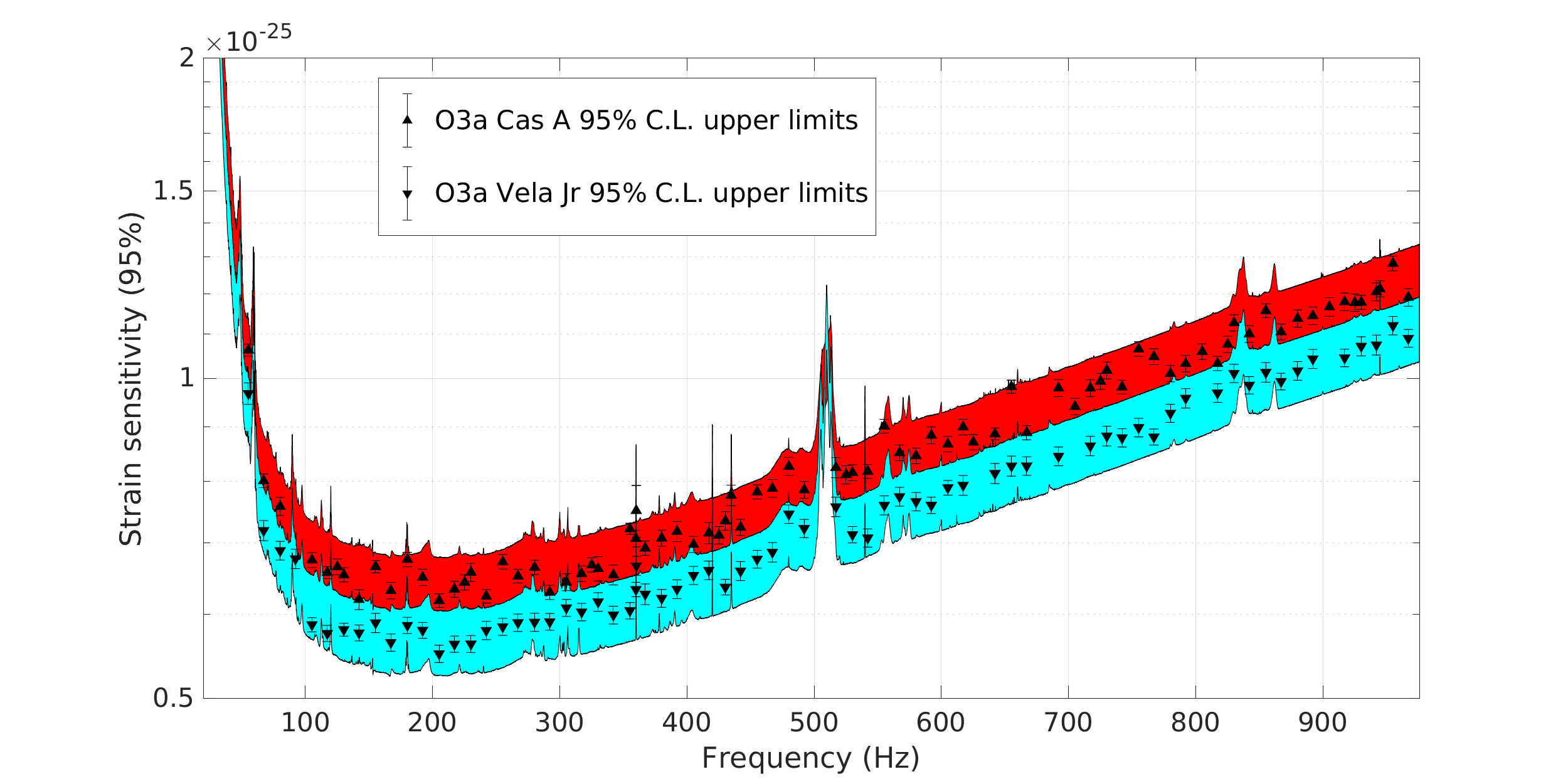}
  \fi
  \caption{{\it Top panel:} Estimated gravitational wave strain amplitude sensitivities (95\%\ efficiency) in each 0.1-Hz sub-band for the
    \casa\ (red band) and \vela\ (cyan band) searches. Conservative uncertainty bands of \comment{$\pm$7\%} are indicated,
    to account for statistical and systematic uncertainties in estimating sensitivity depths, including calibration uncertainties.
    Black triangles (upright -- \casa, inverted -- \vela) denote 0.1-Hz bands for which rigorous upper limits are used to determine estimated
    sensitivity \vs\ frequency.  Sensitivities are estimated for only sub-bands with
    no saturation of the candidate top-list (see Figs.~\ref{fig:CasAoutliercounts}-\ref{fig:VelaJroutliercounts}).
    Sensitivities are based on the absence of any outlier exceeding the frequency-dependent threshold and surviving all stages of
    follow-up, using the sensitivity depths (see Fig.~\ref{fig:sensitivitydepth}) estimated in sample bands
    and rescaled according to the run-average amplitude spectral noise density (H1 and L1 data combined, see Eq.~\ref{eqn:weightedasd}).
    Additional results from prior searches for \casa\ and \vela\ are also shown:
    O1 Einstein@Home 90\%\ C.L. upper limits for \casa\ (magenta curve) and for \vela\ (green curve)~\cite{bib:O1AEI1};
    O3a \casa\ and \vela\ 95\%\ C.L. upper limits using a model-robust Viterbi method (orange curve)~\cite{bib:O3aSNR};
    O3a \vela\ 95\%\ C.L. upper limits using the Band-Sampled-Data directed Frequency Hough method (black curve)~\cite{bib:O3aSNR}.
    The solid red horizontal line indicates the age-based upper limit on \casa\ strain amplitude.
    The dashed (dotted) horizonal blue lines indicate the optimistic (pessimistic) age-based upper limit on \vela\ strain amplitude,
    assuming an age and distance of 700 yr and 0.2 kpc (5100 yr and 1.0 kpc).
    {\it Bottom panel:} Magnification of the sensitivity bands from this analyis over most of the search band ($\sim$40--976 Hz), with 1-$\sigma$ statistical uncertainties shown for
    the individual sparsely sampled upper limits used to estimate the depth.}
  \label{fig:sensitivities}
\end{figure*}

The achieved 95\%-efficiency sensitivities are well below the
age-based strain amplitude limits for these stars over virtually the entire search band of 20--976 Hz.
These sensitivities are shown in Figure~\ref{fig:sensitivities} for both sources
and reach as low as $\sim$\lowesthCasA\ for \casa\ and $\sim$\lowesthVelaJr\ for Vela Jr.
at frequencies near \lowesthfreq\ Hz at 95\%\ efficiency. Conservative uncertainty bands of $\pm$7\%\
are indicated, to account for uncertainties in strain calibration and potential errors in frequency-dependent sensitivity depths.
We have achieved the best sensitivities to date for these sources, reaching 2-3 times below the most sensitive previous
results from the O1 data for \casa.
For \vela\ we reach $\sim$30\% below the most sensitive previous results from the O3 data for frequencies below 600 Hz
and more than 2 times below the most sensitive previous results from the O1 data for \vela\ up to 976 Hz.

\begin{figure*}[htbp]
  \ifshowfigs
  \includegraphics[width=7.5in]{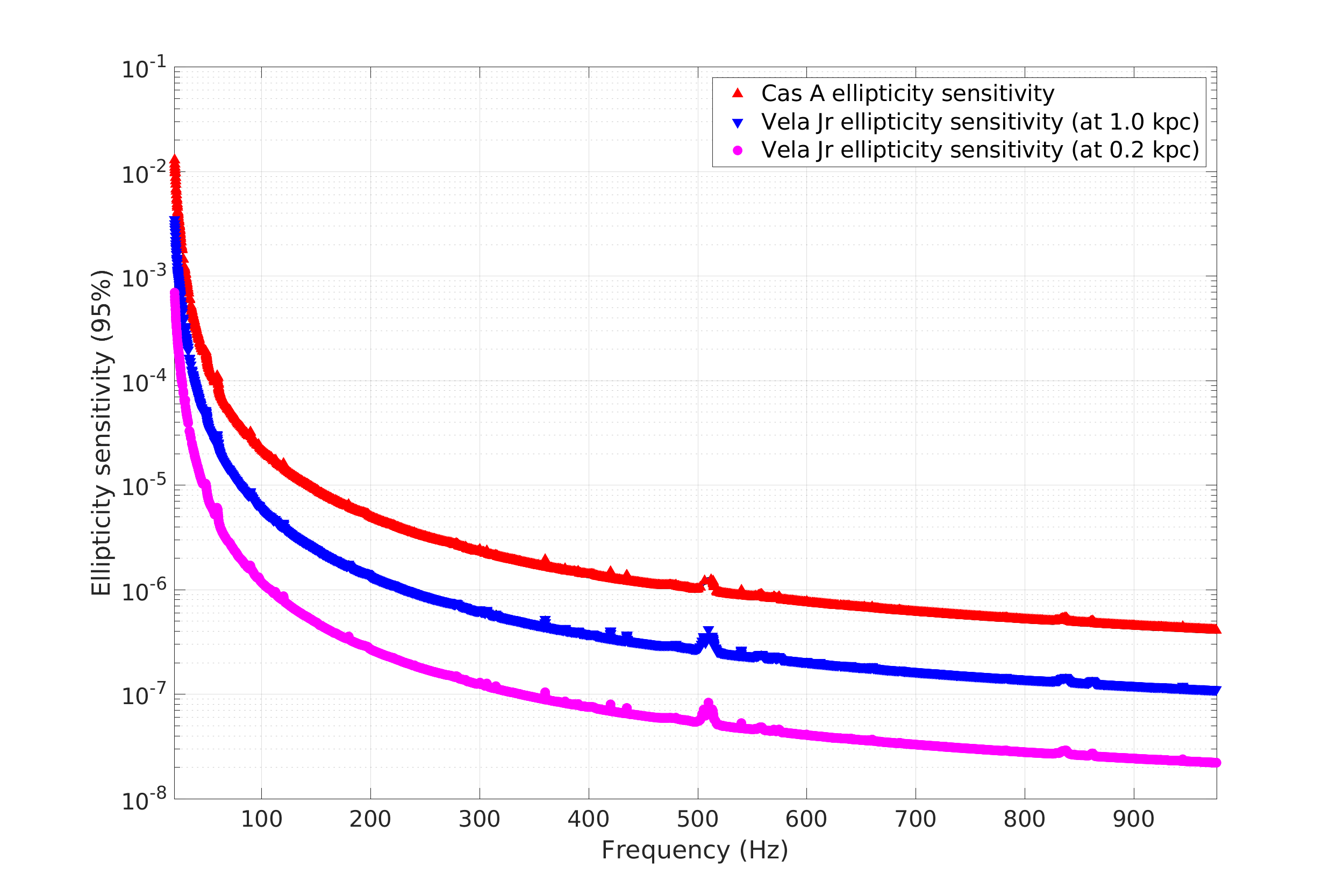}
  \fi
   \caption{Estimated equatorial ellipticity sensitivities (95\%\ efficiency) in each 0.1-Hz sub-band for the \casa\ (red) and \vela\ (blue, magenta) searches,
    derived from the strain amplitude sensitivities shown in Fig.~\ref{fig:sensitivities} assuming a source distance of
    3.3 kpc for \casa, and assuming source distances of 1.0 kpc and 0.2 kpc for \vela\
    (color online).}
  \label{fig:ellipticities}
\end{figure*}

\begin{figure*}[htbp]
  \ifshowfigs
  \includegraphics[width=7.5in]{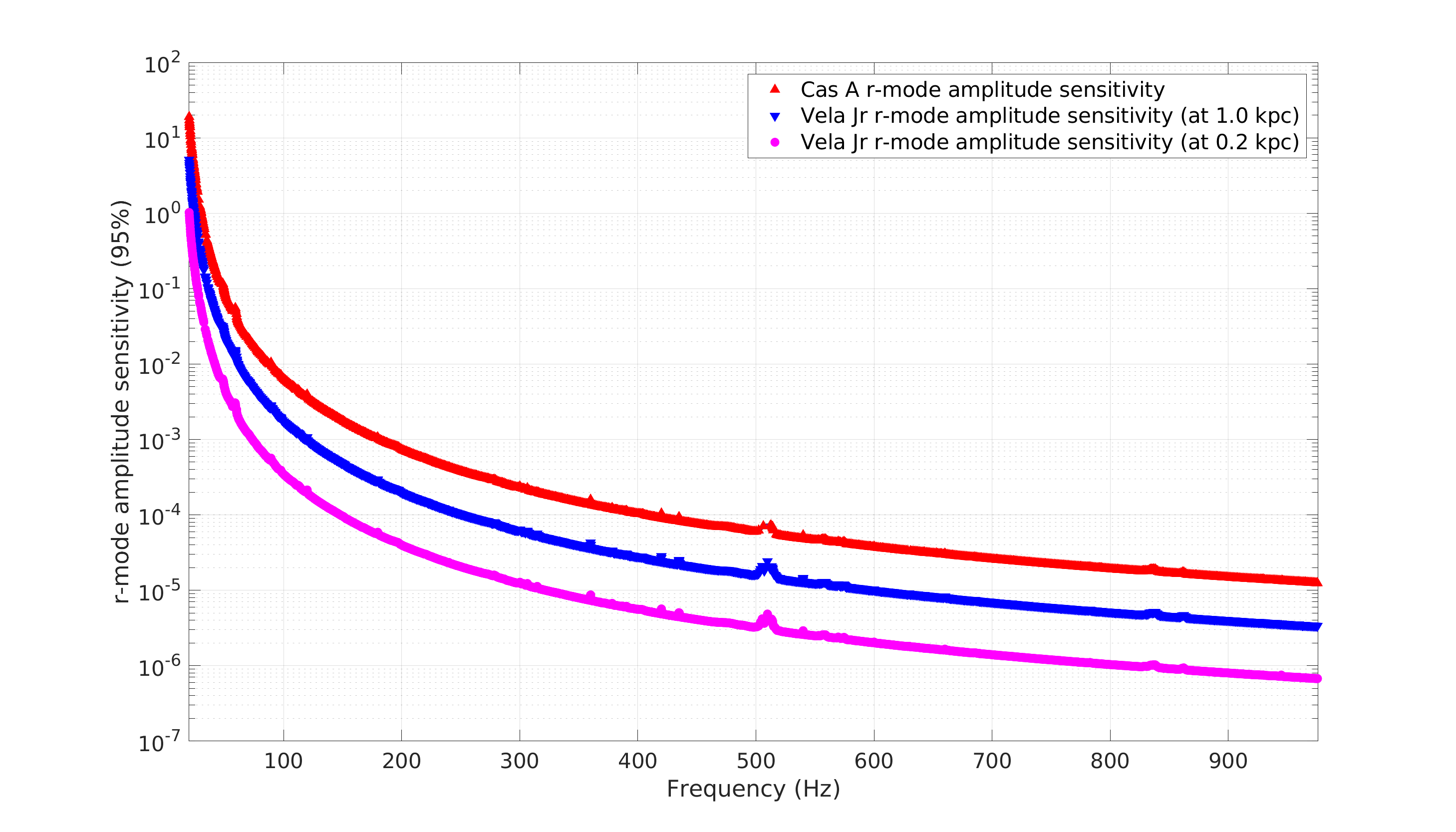}
  \fi
  \caption{Estimated \rmodes\ amplitude $\alpha$ sensitivities (95\%\ efficiency) in each 0.1-Hz sub-band for the \casa\ (red) and \vela\ (blue, magenta) searches,
    derived from the strain amplitude sensitivities shown in Fig.~\ref{fig:sensitivities} assuming a source distance of
    3.3 kpc for \casa, and assuming source distances of 1.0 kpc and 0.2 kpc for \vela\
    (color online).}
  \label{fig:alphas}
\end{figure*}

These sensitivities are translated from strain to equatorial ellipticity $\epsilon$
using Equation~\ref{eqn:ellipticity}, assuming a source distance of 3.3 kpc for \casa, along with both 1.0 kpc and 0.2 kpc for \vela, as shown in 
Figure~\ref{fig:ellipticities}. 
Under an \rmodes\ emission assumption, the strain sensitivities can similarly be translated to \rmode\ amplitude $\alpha$,
shown in Fig.~\ref{fig:alphas}.

As the LIGO, Virgo and KAGRA gravitational wave detectors improve their strain sensitivities in the coming
decade~\cite{bib:scenariospaper}, searches will probe still smaller neutron star deformations, offering 
improved prospects of discovery.

\section{Acknowledgments}
We thank the anonymous journal referee for helpful comments,
expecially concerning the treatment of the third frequency
derivative which led to a refinement of the analysis.
This material is based upon work supported by NSF’s LIGO Laboratory which is a major facility
fully funded by the National Science Foundation.
The authors also gratefully acknowledge the support of
the Science and Technology Facilities Council (STFC) of the
United Kingdom, the Max-Planck-Society (MPS), and the State of
Niedersachsen/Germany for support of the construction of Advanced LIGO 
and construction and operation of the GEO\,600 detector. 
Additional support for Advanced LIGO was provided by the Australian Research Council.
The authors gratefully acknowledge the Italian Istituto Nazionale di Fisica Nucleare (INFN),  
the French Centre National de la Recherche Scientifique (CNRS) and
the Netherlands Organization for Scientific Research (NWO), 
for the construction and operation of the Virgo detector
and the creation and support  of the EGO consortium. 
The authors also gratefully acknowledge research support from these agencies as well as by 
the Council of Scientific and Industrial Research of India, 
the Department of Science and Technology, India,
the Science \& Engineering Research Board (SERB), India,
the Ministry of Human Resource Development, India,
the Spanish Agencia Estatal de Investigaci\'on (AEI),
the Spanish Ministerio de Ciencia e Innovaci\'on and Ministerio de Universidades,
the Conselleria de Fons Europeus, Universitat i Cultura and the Direcci\'o General de Pol\'{\i}tica Universitaria i Recerca del Govern de les Illes Balears,
the Conselleria d'Innovaci\'o, Universitats, Ci\`encia i Societat Digital de la Generalitat Valenciana and
the CERCA Programme Generalitat de Catalunya, Spain,
the National Science Centre of Poland and the European Union – European Regional Development Fund; Foundation for Polish Science (FNP),
the Swiss National Science Foundation (SNSF),
the Russian Foundation for Basic Research, 
the Russian Science Foundation,
the European Commission,
the European Social Funds (ESF),
the European Regional Development Funds (ERDF),
the Royal Society, 
the Scottish Funding Council, 
the Scottish Universities Physics Alliance, 
the Hungarian Scientific Research Fund (OTKA),
the French Lyon Institute of Origins (LIO),
the Belgian Fonds de la Recherche Scientifique (FRS-FNRS), 
Actions de Recherche Concertées (ARC) and
Fonds Wetenschappelijk Onderzoek – Vlaanderen (FWO), Belgium,
the Paris \^{I}le-de-France Region, 
the National Research, Development and Innovation Office Hungary (NKFIH), 
the National Research Foundation of Korea,
the Natural Science and Engineering Research Council Canada,
Canadian Foundation for Innovation (CFI),
the Brazilian Ministry of Science, Technology, and Innovations,
the International Center for Theoretical Physics South American Institute for Fundamental Research (ICTP-SAIFR), 
the Research Grants Council of Hong Kong,
the National Natural Science Foundation of China (NSFC),
the Leverhulme Trust, 
the Research Corporation, 
the Ministry of Science and Technology (MOST), Taiwan,
the United States Department of Energy,
and
the Kavli Foundation.
The authors gratefully acknowledge the support of the NSF, STFC, INFN and CNRS for provision of computational resources.

This document has been assigned LIGO Laboratory document number \texttt{LIGO-P2100298-v8}.

\appendix
\section*{Appendix: Saturated sub-bands}
\label{sec:appendix}

As noted above, some frequency bands were so badly contaminated by instrumental lines that one or more candidate top-lists
from $\fdot$ sub-ranges are saturated ($\ge$1000 candidates) in the initial search.
All 0.1-Hz bands with saturation for the two sources searched are listed in a consolidated format
in Tables~\ref{tab:saturatedbandsCasA}--\ref{tab:saturatedbandsVelaJr} and were visually
examined to verify substantial instrumental contamination.
We do not claim sensitivity to signals in these bands, which sum
for \casa\ (\vela) to 51.0 (40.9) Hz over the full search range of 20--976 Hz.

\begin{table*}[htb]
  \begin{center}
    \def\mc#1{\multicolumn{1}{c}{#1}}
\begin{tabular}{rccrccrccrccrccrccrccrccrccrc}\hline
\mc{$f_{\rm low}$} & \mc{$\Delta f$} &  &\mc{$f_{\rm low}$} & \mc{$\Delta f$} &  &\mc{$f_{\rm low}$} & \mc{$\Delta f$} &  &\mc{$f_{\rm low}$} & \mc{$\Delta f$} &  &\mc{$f_{\rm low}$} & \mc{$\Delta f$} &  &\mc{$f_{\rm low}$} & \mc{$\Delta f$} &  &\mc{$f_{\rm low}$} & \mc{$\Delta f$} &  &\mc{$f_{\rm low}$} & \mc{$\Delta f$} &  &\mc{$f_{\rm low}$} & \mc{$\Delta f$} &  &\mc{$f_{\rm low}$} & \mc{$\Delta f$} \\ 
\mc{(Hz)} & \mc{(Hz)} & & \mc{(Hz)} & \mc{(Hz)} & & \mc{(Hz)} & \mc{(Hz)} & & \mc{(Hz)} & \mc{(Hz)} & & \mc{(Hz)} & \mc{(Hz)} & & \mc{(Hz)} & \mc{(Hz)} & & \mc{(Hz)} & \mc{(Hz)} & & \mc{(Hz)} & \mc{(Hz)} & & \mc{(Hz)} & \mc{(Hz)} & & \mc{(Hz)} & \mc{(Hz)} \\ 
\hline
\hline
20.1 & 0.1
 & \hskip0.036in&\hskip0.036in28.9 & 0.1
 & \hskip0.036in&\hskip0.036in36.9 & 0.2
 & \hskip0.036in&\hskip0.036in45.5 & 0.2
 & \hskip0.036in&\hskip0.036in56.6 & 0.1
 & \hskip0.036in&\hskip0.036in70.0 & 0.2
 & \hskip0.036in&\hskip0.036in119.8 & 0.1
 & \hskip0.036in&\hskip0.036in339.7 & 0.2
 & \hskip0.036in&\hskip0.036in515.0 & 2.2
 & \hskip0.036in&\hskip0.036in638.3 & 0.1
\\
20.7 & 0.2
 & \hskip0.036in&\hskip0.036in29.5 & 0.1
 & \hskip0.036in&\hskip0.036in37.4 & 0.1
 & \hskip0.036in&\hskip0.036in46.0 & 0.2
 & \hskip0.036in&\hskip0.036in56.9 & 0.1
 & \hskip0.036in&\hskip0.036in70.9 & 0.1
 & \hskip0.036in&\hskip0.036in121.0 & 0.1
 & \hskip0.036in&\hskip0.036in348.5 & 0.1
 & \hskip0.036in&\hskip0.036in520.4 & 0.1
 & \hskip0.036in&\hskip0.036in640.4 & 0.1
\\
21.2 & 0.1
 & \hskip0.036in&\hskip0.036in29.8 & 0.2
 & \hskip0.036in&\hskip0.036in38.3 & 0.2
 & \hskip0.036in&\hskip0.036in46.9 & 0.1
 & \hskip0.036in&\hskip0.036in57.5 & 0.1
 & \hskip0.036in&\hskip0.036in71.7 & 0.1
 & \hskip0.036in&\hskip0.036in128.5 & 0.1
 & \hskip0.036in&\hskip0.036in399.9 & 0.1
 & \hskip0.036in&\hskip0.036in520.7 & 0.1
 & \hskip0.036in&\hskip0.036in651.1 & 0.1
\\
21.4 & 0.1
 & \hskip0.036in&\hskip0.036in30.2 & 0.2
 & \hskip0.036in&\hskip0.036in38.7 & 0.1
 & \hskip0.036in&\hskip0.036in47.9 & 0.1
 & \hskip0.036in&\hskip0.036in57.9 & 0.1
 & \hskip0.036in&\hskip0.036in72.1 & 0.1
 & \hskip0.036in&\hskip0.036in130.9 & 0.1
 & \hskip0.036in&\hskip0.036in485.2 & 0.1
 & \hskip0.036in&\hskip0.036in521.4 & 0.2
 & \hskip0.036in&\hskip0.036in652.7 & 0.1
\\
21.8 & 0.2
 & \hskip0.036in&\hskip0.036in30.5 & 0.1
 & \hskip0.036in&\hskip0.036in38.9 & 0.1
 & \hskip0.036in&\hskip0.036in48.9 & 0.1
 & \hskip0.036in&\hskip0.036in58.9 & 0.1
 & \hskip0.036in&\hskip0.036in72.5 & 0.2
 & \hskip0.036in&\hskip0.036in140.2 & 0.1
 & \hskip0.036in&\hskip0.036in487.4 & 0.2
 & \hskip0.036in&\hskip0.036in522.6 & 0.2
 & \hskip0.036in&\hskip0.036in898.6 & 0.2
\\
22.3 & 0.1
 & \hskip0.036in&\hskip0.036in31.1 & 0.1
 & \hskip0.036in&\hskip0.036in39.4 & 0.1
 & \hskip0.036in&\hskip0.036in49.9 & 0.1
 & \hskip0.036in&\hskip0.036in59.4 & 0.3
 & \hskip0.036in&\hskip0.036in73.3 & 0.1
 & \hskip0.036in&\hskip0.036in145.3 & 0.1
 & \hskip0.036in&\hskip0.036in487.9 & 0.1
 & \hskip0.036in&\hskip0.036in525.7 & 0.1
 & \hskip0.036in&\hskip0.036in898.9 & 0.4
\\
22.7 & 0.2
 & \hskip0.036in&\hskip0.036in31.4 & 0.2
 & \hskip0.036in&\hskip0.036in39.7 & 0.1
 & \hskip0.036in&\hskip0.036in50.9 & 0.2
 & \hskip0.036in&\hskip0.036in59.9 & 0.2
 & \hskip0.036in&\hskip0.036in79.7 & 0.1
 & \hskip0.036in&\hskip0.036in151.7 & 0.2
 & \hskip0.036in&\hskip0.036in492.5 & 0.1
 & \hskip0.036in&\hskip0.036in526.3 & 0.1
 & \hskip0.036in&\hskip0.036in906.6 & 0.3
\\
23.5 & 0.2
 & \hskip0.036in&\hskip0.036in31.7 & 0.4
 & \hskip0.036in&\hskip0.036in39.9 & 0.1
 & \hskip0.036in&\hskip0.036in51.2 & 0.1
 & \hskip0.036in&\hskip0.036in62.4 & 0.1
 & \hskip0.036in&\hskip0.036in80.0 & 0.1
 & \hskip0.036in&\hskip0.036in199.9 & 0.1
 & \hskip0.036in&\hskip0.036in493.0 & 0.2
 & \hskip0.036in&\hskip0.036in527.3 & 0.1
 & \hskip0.036in&\hskip0.036in909.8 & 0.4
\\
23.9 & 0.1
 & \hskip0.036in&\hskip0.036in32.3 & 0.1
 & \hskip0.036in&\hskip0.036in40.3 & 0.2
 & \hskip0.036in&\hskip0.036in51.7 & 0.3
 & \hskip0.036in&\hskip0.036in62.8 & 0.1
 & \hskip0.036in&\hskip0.036in83.2 & 0.4
 & \hskip0.036in&\hskip0.036in213.2 & 0.1
 & \hskip0.036in&\hskip0.036in494.7 & 0.1
 & \hskip0.036in&\hskip0.036in527.9 & 0.1
 & \hskip0.036in&\hskip0.036in918.5 & 0.4
\\
24.1 & 0.2
 & \hskip0.036in&\hskip0.036in32.5 & 0.3
 & \hskip0.036in&\hskip0.036in40.6 & 0.1
 & \hskip0.036in&\hskip0.036in52.3 & 0.1
 & \hskip0.036in&\hskip0.036in63.6 & 0.1
 & \hskip0.036in&\hskip0.036in85.1 & 0.1
 & \hskip0.036in&\hskip0.036in246.2 & 0.1
 & \hskip0.036in&\hskip0.036in495.1 & 0.1
 & \hskip0.036in&\hskip0.036in528.2 & 0.2
 & \hskip0.036in&\hskip0.036in922.5 & 0.2
\\
24.6 & 0.1
 & \hskip0.036in&\hskip0.036in32.9 & 0.7
 & \hskip0.036in&\hskip0.036in40.8 & 0.2
 & \hskip0.036in&\hskip0.036in52.5 & 0.2
 & \hskip0.036in&\hskip0.036in63.9 & 0.1
 & \hskip0.036in&\hskip0.036in85.6 & 0.4
 & \hskip0.036in&\hskip0.036in299.2 & 0.8
 & \hskip0.036in&\hskip0.036in495.3 & 0.1
 & \hskip0.036in&\hskip0.036in528.5 & 0.1
 & \hskip0.036in&\hskip0.036in945.1 & 0.2
\\
25.6 & 0.1
 & \hskip0.036in&\hskip0.036in33.9 & 1.1
 & \hskip0.036in&\hskip0.036in41.6 & 0.1
 & \hskip0.036in&\hskip0.036in52.9 & 0.1
 & \hskip0.036in&\hskip0.036in64.2 & 0.3
 & \hskip0.036in&\hskip0.036in87.9 & 0.2
 & \hskip0.036in&\hskip0.036in301.9 & 0.6
 & \hskip0.036in&\hskip0.036in495.9 & 0.1
 & \hskip0.036in&\hskip0.036in533.3 & 0.2
 & \hskip0.036in&\hskip0.036in945.5 & 0.2
\\
25.9 & 0.3
 & \hskip0.036in&\hskip0.036in35.1 & 0.1
 & \hskip0.036in&\hskip0.036in41.8 & 0.2
 & \hskip0.036in&\hskip0.036in53.3 & 0.2
 & \hskip0.036in&\hskip0.036in65.8 & 0.2
 & \hskip0.036in&\hskip0.036in89.9 & 0.1
 & \hskip0.036in&\hskip0.036in303.0 & 0.6
 & \hskip0.036in&\hskip0.036in499.8 & 0.3
 & \hskip0.036in&\hskip0.036in598.6 & 1.3
 & \hskip0.036in&\hskip0.036in\\
26.3 & 0.1
 & \hskip0.036in&\hskip0.036in35.3 & 0.2
 & \hskip0.036in&\hskip0.036in42.4 & 0.1
 & \hskip0.036in&\hskip0.036in53.7 & 0.1
 & \hskip0.036in&\hskip0.036in66.6 & 0.1
 & \hskip0.036in&\hskip0.036in91.1 & 0.1
 & \hskip0.036in&\hskip0.036in305.8 & 0.6
 & \hskip0.036in&\hskip0.036in500.9 & 1.3
 & \hskip0.036in&\hskip0.036in604.0 & 0.9
 & \hskip0.036in&\hskip0.036in\\
26.6 & 0.1
 & \hskip0.036in&\hskip0.036in35.6 & 0.2
 & \hskip0.036in&\hskip0.036in42.8 & 0.3
 & \hskip0.036in&\hskip0.036in54.2 & 0.1
 & \hskip0.036in&\hskip0.036in66.8 & 0.2
 & \hskip0.036in&\hskip0.036in95.8 & 0.2
 & \hskip0.036in&\hskip0.036in307.1 & 0.7
 & \hskip0.036in&\hskip0.036in502.4 & 3.8
 & \hskip0.036in&\hskip0.036in606.1 & 0.9
 & \hskip0.036in&\hskip0.036in\\
26.9 & 0.1
 & \hskip0.036in&\hskip0.036in35.9 & 0.1
 & \hskip0.036in&\hskip0.036in43.6 & 0.4
 & \hskip0.036in&\hskip0.036in54.9 & 0.1
 & \hskip0.036in&\hskip0.036in67.6 & 0.1
 & \hskip0.036in&\hskip0.036in99.8 & 0.3
 & \hskip0.036in&\hskip0.036in314.6 & 0.9
 & \hskip0.036in&\hskip0.036in506.4 & 0.4
 & \hskip0.036in&\hskip0.036in612.0 & 0.8
 & \hskip0.036in&\hskip0.036in\\
27.2 & 0.8
 & \hskip0.036in&\hskip0.036in36.2 & 0.3
 & \hskip0.036in&\hskip0.036in44.4 & 0.3
 & \hskip0.036in&\hskip0.036in55.6 & 0.1
 & \hskip0.036in&\hskip0.036in68.3 & 0.2
 & \hskip0.036in&\hskip0.036in104.3 & 0.1
 & \hskip0.036in&\hskip0.036in323.9 & 0.1
 & \hskip0.036in&\hskip0.036in506.9 & 5.2
 & \hskip0.036in&\hskip0.036in614.5 & 1.0
 & \hskip0.036in&\hskip0.036in\\
28.1 & 0.7
 & \hskip0.036in&\hskip0.036in36.6 & 0.1
 & \hskip0.036in&\hskip0.036in44.9 & 0.1
 & \hskip0.036in&\hskip0.036in55.9 & 0.1
 & \hskip0.036in&\hskip0.036in69.2 & 0.1
 & \hskip0.036in&\hskip0.036in107.1 & 0.1
 & \hskip0.036in&\hskip0.036in333.2 & 0.2
 & \hskip0.036in&\hskip0.036in512.9 & 1.1
 & \hskip0.036in&\hskip0.036in629.7 & 1.0
 & \hskip0.036in&\hskip0.036in\\
\hline
\end{tabular}

    \caption{Frequency bands with saturation in the first stage of the \casa\ search ($\ge$1000 outliers above threshold in a 0.1-Hz band for at
      least one sub-range of frequency derivatives. Each pair of numbers gives the lower limit of frequency and the width of the band
      affected. Consecutive 0.1-Hz bands are concatenated for compactness. These bands are excluded from the \casa\ sensitivity curve shown
      in Fig.~\ref{fig:sensitivities}}
\label{tab:saturatedbandsCasA}
\end{center}
\end{table*}

\begin{table*}[htb]
  \begin{center}
    \def\mc#1{\multicolumn{1}{c}{#1}}
\begin{tabular}{rccrccrccrccrccrccrccrccrccrc}\hline
\mc{$f_{\rm low}$} & \mc{$\Delta f$} &  &\mc{$f_{\rm low}$} & \mc{$\Delta f$} &  &\mc{$f_{\rm low}$} & \mc{$\Delta f$} &  &\mc{$f_{\rm low}$} & \mc{$\Delta f$} &  &\mc{$f_{\rm low}$} & \mc{$\Delta f$} &  &\mc{$f_{\rm low}$} & \mc{$\Delta f$} &  &\mc{$f_{\rm low}$} & \mc{$\Delta f$} &  &\mc{$f_{\rm low}$} & \mc{$\Delta f$} &  &\mc{$f_{\rm low}$} & \mc{$\Delta f$} &  &\mc{$f_{\rm low}$} & \mc{$\Delta f$} \\ 
\mc{(Hz)} & \mc{(Hz)} & & \mc{(Hz)} & \mc{(Hz)} & & \mc{(Hz)} & \mc{(Hz)} & & \mc{(Hz)} & \mc{(Hz)} & & \mc{(Hz)} & \mc{(Hz)} & & \mc{(Hz)} & \mc{(Hz)} & & \mc{(Hz)} & \mc{(Hz)} & & \mc{(Hz)} & \mc{(Hz)} & & \mc{(Hz)} & \mc{(Hz)} & & \mc{(Hz)} & \mc{(Hz)} \\ 
\hline
\hline
20.1 & 0.1
 & \hskip0.036in&\hskip0.036in27.3 & 0.6
 & \hskip0.036in&\hskip0.036in36.0 & 0.1
 & \hskip0.036in&\hskip0.036in43.6 & 0.2
 & \hskip0.036in&\hskip0.036in56.6 & 0.1
 & \hskip0.036in&\hskip0.036in79.7 & 0.1
 & \hskip0.036in&\hskip0.036in303.1 & 0.4
 & \hskip0.036in&\hskip0.036in503.5 & 0.4
 & \hskip0.036in&\hskip0.036in524.6 & 0.1
 & \hskip0.036in&\hskip0.036in652.8 & 0.1
\\
20.5 & 0.1
 & \hskip0.036in&\hskip0.036in28.0 & 0.1
 & \hskip0.036in&\hskip0.036in36.3 & 0.4
 & \hskip0.036in&\hskip0.036in44.0 & 0.1
 & \hskip0.036in&\hskip0.036in56.9 & 0.1
 & \hskip0.036in&\hskip0.036in80.0 & 0.1
 & \hskip0.036in&\hskip0.036in305.9 & 0.5
 & \hskip0.036in&\hskip0.036in504.0 & 1.3
 & \hskip0.036in&\hskip0.036in525.2 & 0.1
 & \hskip0.036in&\hskip0.036in666.6 & 0.1
\\
20.7 & 0.1
 & \hskip0.036in&\hskip0.036in28.2 & 0.6
 & \hskip0.036in&\hskip0.036in36.8 & 0.1
 & \hskip0.036in&\hskip0.036in44.5 & 0.2
 & \hskip0.036in&\hskip0.036in57.5 & 0.1
 & \hskip0.036in&\hskip0.036in83.2 & 0.3
 & \hskip0.036in&\hskip0.036in307.2 & 0.5
 & \hskip0.036in&\hskip0.036in505.5 & 0.4
 & \hskip0.036in&\hskip0.036in525.8 & 0.1
 & \hskip0.036in&\hskip0.036in898.7 & 0.2
\\
21.4 & 0.2
 & \hskip0.036in&\hskip0.036in29.0 & 0.1
 & \hskip0.036in&\hskip0.036in37.0 & 0.1
 & \hskip0.036in&\hskip0.036in45.5 & 0.1
 & \hskip0.036in&\hskip0.036in58.0 & 0.1
 & \hskip0.036in&\hskip0.036in85.7 & 0.3
 & \hskip0.036in&\hskip0.036in314.8 & 0.5
 & \hskip0.036in&\hskip0.036in506.5 & 0.3
 & \hskip0.036in&\hskip0.036in526.3 & 0.2
 & \hskip0.036in&\hskip0.036in899.1 & 0.4
\\
21.8 & 0.2
 & \hskip0.036in&\hskip0.036in29.5 & 0.1
 & \hskip0.036in&\hskip0.036in37.4 & 0.1
 & \hskip0.036in&\hskip0.036in46.0 & 0.2
 & \hskip0.036in&\hskip0.036in59.0 & 0.1
 & \hskip0.036in&\hskip0.036in87.9 & 0.2
 & \hskip0.036in&\hskip0.036in333.3 & 0.2
 & \hskip0.036in&\hskip0.036in507.2 & 0.5
 & \hskip0.036in&\hskip0.036in527.1 & 0.1
 & \hskip0.036in&\hskip0.036in906.8 & 0.2
\\
22.3 & 0.1
 & \hskip0.036in&\hskip0.036in29.9 & 0.2
 & \hskip0.036in&\hskip0.036in38.3 & 0.2
 & \hskip0.036in&\hskip0.036in46.5 & 0.1
 & \hskip0.036in&\hskip0.036in59.4 & 0.3
 & \hskip0.036in&\hskip0.036in89.9 & 0.1
 & \hskip0.036in&\hskip0.036in400.0 & 0.1
 & \hskip0.036in&\hskip0.036in507.9 & 1.8
 & \hskip0.036in&\hskip0.036in527.8 & 0.1
 & \hskip0.036in&\hskip0.036in910.0 & 0.3
\\
22.5 & 0.1
 & \hskip0.036in&\hskip0.036in30.2 & 0.2
 & \hskip0.036in&\hskip0.036in38.7 & 0.1
 & \hskip0.036in&\hskip0.036in48.0 & 0.1
 & \hskip0.036in&\hskip0.036in59.9 & 0.2
 & \hskip0.036in&\hskip0.036in91.1 & 0.1
 & \hskip0.036in&\hskip0.036in485.2 & 0.1
 & \hskip0.036in&\hskip0.036in510.0 & 1.9
 & \hskip0.036in&\hskip0.036in528.3 & 0.2
 & \hskip0.036in&\hskip0.036in918.7 & 0.3
\\
22.7 & 0.1
 & \hskip0.036in&\hskip0.036in30.5 & 0.1
 & \hskip0.036in&\hskip0.036in39.7 & 0.1
 & \hskip0.036in&\hskip0.036in50.0 & 0.1
 & \hskip0.036in&\hskip0.036in62.4 & 0.1
 & \hskip0.036in&\hskip0.036in95.8 & 0.2
 & \hskip0.036in&\hskip0.036in485.7 & 0.1
 & \hskip0.036in&\hskip0.036in513.2 & 0.6
 & \hskip0.036in&\hskip0.036in533.3 & 0.2
 & \hskip0.036in&\hskip0.036in922.6 & 0.2
\\
23.5 & 0.1
 & \hskip0.036in&\hskip0.036in30.9 & 0.1
 & \hskip0.036in&\hskip0.036in40.0 & 0.1
 & \hskip0.036in&\hskip0.036in50.9 & 0.2
 & \hskip0.036in&\hskip0.036in62.8 & 0.1
 & \hskip0.036in&\hskip0.036in99.9 & 0.2
 & \hskip0.036in&\hskip0.036in486.4 & 0.1
 & \hskip0.036in&\hskip0.036in515.2 & 1.0
 & \hskip0.036in&\hskip0.036in598.9 & 0.8
 & \hskip0.036in&\hskip0.036in945.2 & 0.2
\\
24.0 & 0.1
 & \hskip0.036in&\hskip0.036in31.2 & 0.1
 & \hskip0.036in&\hskip0.036in40.3 & 0.2
 & \hskip0.036in&\hskip0.036in51.6 & 0.3
 & \hskip0.036in&\hskip0.036in63.6 & 0.1
 & \hskip0.036in&\hskip0.036in107.1 & 0.1
 & \hskip0.036in&\hskip0.036in487.2 & 0.2
 & \hskip0.036in&\hskip0.036in516.5 & 0.5
 & \hskip0.036in&\hskip0.036in604.2 & 0.5
 & \hskip0.036in&\hskip0.036in945.6 & 0.2
\\
24.2 & 0.1
 & \hskip0.036in&\hskip0.036in31.4 & 0.2
 & \hskip0.036in&\hskip0.036in40.6 & 0.1
 & \hskip0.036in&\hskip0.036in52.0 & 0.1
 & \hskip0.036in&\hskip0.036in64.0 & 0.1
 & \hskip0.036in&\hskip0.036in119.8 & 0.1
 & \hskip0.036in&\hskip0.036in492.6 & 0.1
 & \hskip0.036in&\hskip0.036in518.4 & 0.1
 & \hskip0.036in&\hskip0.036in606.4 & 0.5
 & \hskip0.036in&\hskip0.036in\\
24.5 & 0.1
 & \hskip0.036in&\hskip0.036in31.7 & 0.4
 & \hskip0.036in&\hskip0.036in40.8 & 0.3
 & \hskip0.036in&\hskip0.036in52.3 & 0.1
 & \hskip0.036in&\hskip0.036in64.2 & 0.3
 & \hskip0.036in&\hskip0.036in128.5 & 0.1
 & \hskip0.036in&\hskip0.036in493.1 & 0.2
 & \hskip0.036in&\hskip0.036in519.2 & 0.1
 & \hskip0.036in&\hskip0.036in612.3 & 0.4
 & \hskip0.036in&\hskip0.036in\\
25.5 & 0.2
 & \hskip0.036in&\hskip0.036in32.3 & 0.1
 & \hskip0.036in&\hskip0.036in41.6 & 0.1
 & \hskip0.036in&\hskip0.036in52.5 & 0.2
 & \hskip0.036in&\hskip0.036in66.7 & 0.1
 & \hskip0.036in&\hskip0.036in140.2 & 0.1
 & \hskip0.036in&\hskip0.036in493.8 & 0.1
 & \hskip0.036in&\hskip0.036in519.9 & 0.1
 & \hskip0.036in&\hskip0.036in614.7 & 0.6
 & \hskip0.036in&\hskip0.036in\\
26.0 & 0.2
 & \hskip0.036in&\hskip0.036in32.5 & 1.1
 & \hskip0.036in&\hskip0.036in41.8 & 0.1
 & \hskip0.036in&\hskip0.036in53.3 & 0.2
 & \hskip0.036in&\hskip0.036in68.3 & 0.1
 & \hskip0.036in&\hskip0.036in151.7 & 0.2
 & \hskip0.036in&\hskip0.036in495.1 & 0.1
 & \hskip0.036in&\hskip0.036in520.4 & 0.2
 & \hskip0.036in&\hskip0.036in629.8 & 0.8
 & \hskip0.036in&\hskip0.036in\\
26.3 & 0.1
 & \hskip0.036in&\hskip0.036in33.8 & 0.5
 & \hskip0.036in&\hskip0.036in42.0 & 0.1
 & \hskip0.036in&\hskip0.036in53.7 & 0.1
 & \hskip0.036in&\hskip0.036in69.4 & 0.1
 & \hskip0.036in&\hskip0.036in199.9 & 0.2
 & \hskip0.036in&\hskip0.036in499.9 & 0.2
 & \hskip0.036in&\hskip0.036in520.7 & 0.1
 & \hskip0.036in&\hskip0.036in638.3 & 0.1
 & \hskip0.036in&\hskip0.036in\\
26.5 & 0.2
 & \hskip0.036in&\hskip0.036in34.4 & 0.5
 & \hskip0.036in&\hskip0.036in42.4 & 0.2
 & \hskip0.036in&\hskip0.036in55.6 & 0.1
 & \hskip0.036in&\hskip0.036in70.1 & 0.1
 & \hskip0.036in&\hskip0.036in299.3 & 0.8
 & \hskip0.036in&\hskip0.036in501.1 & 0.9
 & \hskip0.036in&\hskip0.036in521.6 & 0.1
 & \hskip0.036in&\hskip0.036in640.5 & 0.1
 & \hskip0.036in&\hskip0.036in\\
26.9 & 0.2
 & \hskip0.036in&\hskip0.036in35.0 & 0.8
 & \hskip0.036in&\hskip0.036in42.8 & 0.3
 & \hskip0.036in&\hskip0.036in56.0 & 0.1
 & \hskip0.036in&\hskip0.036in72.5 & 0.2
 & \hskip0.036in&\hskip0.036in302.1 & 0.3
 & \hskip0.036in&\hskip0.036in502.6 & 0.8
 & \hskip0.036in&\hskip0.036in522.5 & 0.1
 & \hskip0.036in&\hskip0.036in651.1 & 0.1
 & \hskip0.036in&\hskip0.036in\\
\hline
\end{tabular}

    \caption{Frequency bands with saturation in the first stage of the \vela\ search ($\ge$1000 outliers above threshold in a 0.1-Hz band for at
      least one sub-range of frequency derivatives. Each pair of numbers gives the lower limit of frequency and the width of the band
      affected. Consecutive 0.1-Hz bands are concatenated for compactness. These bands are excluded from the \vela\ sensitivity curve shown
      in Fig.~\ref{fig:sensitivities}}
\label{tab:saturatedbandsVelaJr}
\end{center}
\end{table*}

\newif\ifshowrefs\showrefsfalse
\showrefstrue

\ifshowrefs

\fi

\let\author\myauthor
\let\affiliation\myaffiliation
\let\maketitle\mymaketitle
\title{Authors}


\author{R.~Abbott}
\affiliation{LIGO Laboratory, California Institute of Technology, Pasadena, CA 91125, USA}
\author{T.~D.~Abbott}
\affiliation{Louisiana State University, Baton Rouge, LA 70803, USA}
\author{F.~Acernese}
\affiliation{Dipartimento di Farmacia, Universit\`a di Salerno, I-84084 Fisciano, Salerno, Italy}
\affiliation{INFN, Sezione di Napoli, Complesso Universitario di Monte S. Angelo, I-80126 Napoli, Italy}
\author{K.~Ackley}
\affiliation{OzGrav, School of Physics \& Astronomy, Monash University, Clayton 3800, Victoria, Australia}
\author{C.~Adams}
\affiliation{LIGO Livingston Observatory, Livingston, LA 70754, USA}
\author{N.~Adhikari}
\affiliation{University of Wisconsin-Milwaukee, Milwaukee, WI 53201, USA}
\author{R.~X.~Adhikari}
\affiliation{LIGO Laboratory, California Institute of Technology, Pasadena, CA 91125, USA}
\author{V.~B.~Adya}
\affiliation{OzGrav, Australian National University, Canberra, Australian Capital Territory 0200, Australia}
\author{C.~Affeldt}
\affiliation{Max Planck Institute for Gravitational Physics (Albert Einstein Institute), D-30167 Hannover, Germany}
\affiliation{Leibniz Universit\"at Hannover, D-30167 Hannover, Germany}
\author{D.~Agarwal}
\affiliation{Inter-University Centre for Astronomy and Astrophysics, Pune 411007, India}
% ...




\maketitle
\newpage

\end{document}